\documentclass[aps,prb,twocolumn,showpacs,superscriptaddress,groupedaddress]{revtex4-1}
\usepackage{dsfont,amsthm,amsbsy}
\usepackage{amssymb}
\usepackage{amsmath}
\usepackage{graphicx}
\usepackage{epstopdf}
\usepackage{subfigure}
\usepackage{natbib}
\usepackage{epsfig}
\usepackage{amsfonts}
\usepackage{mathrsfs}
\usepackage{sidecap}
\usepackage{lipsum}
\usepackage[toc,page,title,titletoc,header]{appendix}
\usepackage[colorlinks,linkcolor=blue,citecolor=blue,anchorcolor=blue]{hyperref}
 \hypersetup{ hidelinks, }
\usepackage{dsfont,amsthm,amsbsy}
\usepackage{resizegather}
\usepackage{caption}
\usepackage{tikz}
\usepackage{float}
\usepackage[normalem]{ulem}
\usepackage{cancel}
\newcommand{\oi}{{\rm i}}

\newcommand{\bl}{\begin{aligned}}
\newcommand{\el}{\end{aligned}}
\def\be{\begin{equation}}
\def\ee{\end{equation}}

\def\bi{\begin{itemize}}
\def\ei{\end{itemize}}
\def\bn{\begin{enumerate}}
\def\en{\end{enumerate}}
\def\bea{\begin{eqnarray}}
\def\eea{\end{eqnarray}}
\def\no{\nonumber}
\def\ba{\begin{array}}
\def\ea{\end{array}}
\def\bd{\begin{displaymath}}
\def\ed{\end{displaymath}}

\def\ra{\rangle}

\begin{document}

\title{Decoherence from spin environments: \\ Loschmidt echo and quasiparticle excitations}

\author{R. Jafari$^{1,2,3}$}
\email[]{Electronic address: rohollah.jafari@gmail.com \\(corresponding author)}
\author{Henrik Johannesson$^{1,3}$}

\affiliation{$^{1}$Beijing Computational Science Research Center, Beijing 100094, China\\
$^{2}$Department of Physics, Institute for Advanced Studies in Basic Sciences (IASBS), Zanjan 45137-66731, Iran\\
$^{3}$Department of Physics, University of Gothenburg, SE 412 96 Gothenburg, Sweden}


\begin{abstract}
We revisit the problem of decoherence of a qubit centrally coupled to an interacting spin environment, here modeled by a quantum compass chain
or an extended XY model in a staggered magnetic field. These two models both support distinct spin liquid phases, adding a new element to the problem.
By analyzing their Loschmidt echoes when perturbed by the qubit we find that a fast decoherence of the qubit is conditioned on the presence of propagating quasiparticles
which couple to it. Different from expectations based on earlier works on central spin models, our result implies that
the closeness of an environment to a quantum phase transition is neither a sufficient nor a necessary condition for an accelerated decoherence rate of a qubit.
\end{abstract}

\maketitle


\section{Introduction \label{introduction}}

Progress in experiments on cold atoms trapped in optical lattices has made possible studies of interacting quantum many-particle systems at an
unprecedented level of control \cite{Bloch2008}. A problem that may soon be amenable to experimental probes is that of decoherence induced by an environment which is
close to a quantum phase transition (QPT) \cite{Sachdev2011}. Decoherence refers to the process where the entanglement between a system and its environment makes the
system give up quantum information, turning its pure state, when isolated, into a mixed state. The loss of coherence lies at the heart of the measurement problem \cite{Sclosshauer2005}
and that of understanding the ``quantum-to-classical" transition \cite{Zurek2003}, and also presents a major challenge for realizing quantum information protocols \cite{Shor1995}.

Starting with the work of Dobrovitski {\em et al.} \cite{Dobrovitski2003b}, there have been a number of studies modeling decoherence due to an interacting zero-temperature  spin environment  \cite{Melikidze2004,Cucchietti2005,Rossini2007,Ou2007,Liu2007,Hanson2008,Cormick2008,Lai2008,Cheng2009}.
An important contribution was made by Quan {\em et al.} \cite{Quan2006} who addressed the role of quantum criticality of the environment. Inspired by the Hepp-Coleman model \cite{Hepp1972,Bell1975}, Quan {\em et al.} introduced a central spin model where a qubit, or two-level system, is coupled to all spins of a surrounding environment, here taken to be an Ising chain in a transverse magnetic field. Other works soon followed, based on the same type of setup, but with an XY chain \cite{Yuan2007a} representing the interacting spin environment.
The conclusion drawn from these and similar investigations is that the qubit state decoheres faster when the environment approaches a QPT \cite{Quan2006,Yuan2007a,Cucchietti2007,Sun:2007aa,Ma2007,Mostame2007,Rossini2008,Haikka2012,Sharma2012}. The high sensitivity of the ground state of the environment to a perturbation from the qubit when close to a QPT is here believed to be the reason why the time evolution of the entanglement, and by that, the decoherence rate of the qubit, gets accelerated.

A very useful conceptual tool for exploring this circle of problems is that of the Loschmidt echo \cite{Gorin2006} which provides a measure of the stiffness of the environment to the perturbation
from the qubit. The Loschmidt echo in a central spin model coincides with the square
of the decoherence factor of the qubit, being proportional to the rate of decoherence. Thus, a study of the Loschmidt echo allows for a detailed analysis of the decoherence
process. As a case in point, Quan {\em et al.} \cite{Quan2006} referred to the fast initial decay of the Loschmidt echo exactly at the critical point of the transverse field Ising chain to argue that the decoherence of a qubit is accelerated by the criticality of its environment. Further,
Haikka {\em et al.} \cite{Haikka2012} relied on the observation of the monotonic short-time decay of the Loschmidt echo for the same setup to argue that the reduced dynamics of the qubit for this case is purely Markovian \cite{Vega2017}. This would mean that the critical point blocks any backflow of information from the environment into the qubit at short initial times, a flow otherwise expected from the typical appearance of revivals in a Loschmidt echo for a finite system over larger time scales.

Motivated by the prospect of future experiments that may probe the connection between decoherence and quantum criticality, we have revisited the problem of qubit decoherence in an interacting spin environment. However, our aim has not been to propose or analyze a particular experimental arrangement. Rather, we wish to examine
the very notion that the closeness of an environment to a QPT is intrinsically linked to an accelerated decoherence rate of the system to which it couples. Like others before us we shall
take advantage of a central spin model, allowing for a transparent analysis, but now using a generalized quantum compass chain (QCC) \cite{You2014a} and an extended XY model in a staggered magnetic field \cite{Titvinidze2003} as environmental models.

The QCC \cite{You2014a} incorporates a family of one-dimensional (1D) compass models \cite{Brezicki2007,You2008,Eriksson2009,Jafari2011,You2014b,Jafari2016} which serve as ``stripped-down" versions of the more familiar compass models defined for spins on two- or three-dimensional lattices. The common denominator is the structure of their Hamiltonians, being built from directional competing Ising-like interactions between neighboring spin components, with different components interacting on different bonds of the lattice \cite{Nussinov2015}. The QCC exhibits a QPT between two disordered phases with different short-range spin correlations, occurring when the Ising-like interactions are fine tuned to become isotropic \cite{You2014a}. The extended XY model that we shall also consider as a description of a possible spin environment is typified by the presence of three-site XY spin interactions \cite{Titvinidze2003}. This model also exhibits distinct ground state phases, but of a different character from the QCC, accessible by tuning the spin couplings and/or a uniform or staggered magnetic field.

Both models feature spectra with structures that add a level of complexity beyond that of the simpler models hitherto considered in the literature. In particular, the presence of distinct spin liquid phases separated by QPTs brings a new element to the problem. This property, together with the fact that the models are exactly solvable, is the reason why we have selected them for our study.  By a detailed analysis, based on the exact solution of the respective model \cite{You2014a,Titvinidze2003}, we arrive at the conclusion that the notion of an intrinsic connection between quantum phase transitions and strong decoherence misses out on the very mechanism which drives an accelerated decoherence rate: What matters is not the presence of a quantum phase transition {\em per se}, but instead the availability of propagating quasiparticles which couple to the qubit via a back action (as signaled by their having an impact on the Loschmidt echo). Such quasiparticles may indeed be expected to appear at a QPT, but as our case study of the QCC reveals, this is not necessarily so. As transpires when taking the extended XY model as environmental model, quasiparticles of this type may instead appear in a stable massless phase away from a QPT. These results bring new light on how to understand the enhanced decoherence experienced by systems coupled to an interacting environment.

The paper is organized as follows: In the next section we define the central spin model with the QCC as environment, and we also provide some background material. In Sec. III we
diagonalize the QCC Hamiltonian to obtain its spectrum and eigenstates, and from this, we construct exact expressions for the Loschmidt echo. Numerical case studies of the Loschmidt echo point to
the crucial role of quasiparticle excitations in the decoherence process, and to this we add supporting evidence by a theoretical analysis. This section is divided into two parts, treating the QCC without and with a magnetic field respectively. In Sec. IV we carry out essentially the same type of analysis as in Sec. III, but now with the extended XY model in a transverse field as environmental model.
Sec. V, finally, contains a brief summary and discussion. Some technical details are placed in the Appendix.

%
%
\section{Decoherence of a qubit coupled to a quantum compass chain \label{proposal}}

Following the original proposal by Quan {\it et al.} \cite{Quan2006} for modeling the decoherence of a qubit coupled to an interacting spin environment,
we consider a composite system in a factorized pure state at time $t=0$,
\bea
\label{initial}
|\psi(0)\rangle =|\phi_q(0)\ra \!\otimes |\phi_{\text{env}}(0)\ra.
\eea
Here
$|\phi_q(0)\ra = c_g|g\ra + c_e|e\ra$, with $|c_g|^2+|c_e|^2=1$, is the qubit state, while $|\phi(0)_{\text{env}}\rangle$ is the ground state of the environment when isolated.
We take the environment to be an $N$-site QCC in a transverse field $h$, with Hamiltonian  \cite{You2014a}
\bea
\label{QCChamiltonian}
\no
H_{\text{env}}&\!=&\!\!\sum_{n=1}^{N/2}\!\left(J_{o}\tilde{\sigma}_{2n-1}^{+}\tilde{\sigma}_{2n}^{+}
\!+\!J_{e}\tilde{\sigma}_{2n}^{-}\tilde{\sigma}_{2n+1}^{-}\!-\!h(\sigma^{z}_{2n-1}\!+\!\sigma^{z}_{2n})\right),\\
\eea
satisfying periodic boundary conditions. The exchange couplings $J_{o}$ and $J_{e}$ are defined on ``odd" $(2n-1,2n)$ and ``even" $(2n,2n+1)$ lattice bonds respectively, and $\tilde{\sigma}_{2i}^{\pm}$ are pseudospin operators constructed as linear combinations of the Pauli matrices $\sigma^{x}$ and $\sigma^{y}$, $\tilde{\sigma}_{i}^{\pm}=\cos\theta\, \sigma^{x}_{i} \pm\sin\theta\, \sigma^{y}_{i}$.

Denoting the energy difference between the ground state $|g\rangle$ and the excited state $|e\rangle$ of the qubit by $\omega_e$, its free Hamiltonian can be written
\bea
\label{qubit}
H_{\text{q}}&=&\omega_{e}|e\rangle\langle e|.
\eea
The qubit is assumed to couple with equal strength $\delta$ to all pseudospins,
\bea
H_{\text{int}}=-\delta|e\rangle\langle e|\sum_{n=1}^{N} \sigma^{z}_{n},
\eea
with $\delta \ll \omega_e$, making the full Hamiltonian
\bea
\label{totalH}
H=H_{\text{env}}+H_{\text{q}} +H_{\text {int}}
\eea
belong to the class of central spin models first conceived by Gaudin \cite{Gaudin1976}. The new element in our setup is that we have substituted the QCC in (\ref{QCChamiltonian}) for the time-honored use of the quantum Ising \cite{Quan2006} or XY \cite{Yuan2007a} chains. While the problem by this becomes rather more complex, we shall find that it is still amenable to an exact analysis.

Noting that $[H_{\text{q}}, H_{\text{int}}] = 0$, it is easy to verify that the time evolution of the composite state in Eq. (\ref{initial})
splits into two terms:
\bea
\label{eq1}
|\psi(t)\rangle&=&c_{g}|g\rangle\otimes \exp(-iH_{\text{env}}t)\, |\phi_{\text{env}}(0)\rangle \no \\
&+&e^{-i \omega_{e}t}c_{e}|e\rangle\otimes \exp(-iH^{(\delta)}_{\text{env}}t)\,|\phi_{\text{env}}(0)\rangle.
\eea
The first term
evolves with the unperturbed Hamiltonian of the environment, $H_{\text{env}}$, while in the second term the state of the environment evolves with
\begin{equation}
\label{pertqcc}
H^{(\delta)}_{\text{env}}= H_{\text{env}} + V^{(\delta)}_{\text{env}},
\end{equation}
 where $V^{(\delta)}_{\text{env}}=-\delta\sum_{n=1}^{N}\sigma_{n}^{z}$ is an effective potential from the coupling to the qubit,
leading to a redefinition of the magnetic field, $h \rightarrow h+\delta$.
As expected from the vanishing of the commutator $[H_{\text{q}}, H_{\text{int}}]$, the form of the time evolution in (\ref{eq1}) manifests a pure decoherence of the qubit, with no
exchange of energy with the environment.

Tracing out the states of the environment, the time-evolved reduced density matrix of the qubit can be written as
\begin{eqnarray}
\label{qubitRho} \no
\rho_q(t) &=& |c_g|^2|g \rangle \langle g| + |c_e|^2|e \rangle \langle e| \\
&+& e^{-i\omega_et}c^{\ast}_g c_e  \nu(t) |e\rangle \langle g| +  e^{i\omega_et}c_g c^{\ast}_e \nu^{\ast}(t) |g\rangle \langle e|,
\end{eqnarray}
where the decoherence factor $\nu(t)$ $-$ quantifying how the qubit state decoheres with time $-$ takes the form
\bea
\nu(t) \!=\! \langle \phi_{\text{env}}(0)| \exp(iH_{\text{env}}^{(\delta)}t) \exp(-iH_{\text{env}}t) |\phi_{\text{env}}(0)\rangle,
\eea
given our assumption that the initial state of the environment is pure, $\rho_{\text{env}}(0) = |\phi_{\text{env}}(0)\rangle \langle \phi_{\text{env}}(0)|$.
The absolute square of the decoherence factor $\nu(t)$ equals the Loschmidt echo (LE) of the environment,
\begin{eqnarray}
\label{Loschmidt}
{\cal L}(t) &\!=\!&  |\nu(t)|^2   \\
&\!=\!& |\langle \phi_{\text{env}}(0)|\exp(iH_{\text{env}}^{(\delta)}t) \exp(-iH_{\text{env}}t) |\phi_{\text{env}}(0)\rangle|^2. \nonumber
\end{eqnarray}
As expressed by this equation, an LE \cite{Gorin2006} quantifies the overlap between two states at time $t$ evolved from the same initial state but with different Hamiltonians, the one differing from the other by a small perturbation.  As such it provides a measure of the robustness of the time evolution of a system when subject to small perturbations. The simple relation between $\nu(t)$ and ${\cal L}(t)$ in Eq. (\ref{Loschmidt}) formalizes our intuition that an environment whose time evolution is highly sensitive to a perturbation will also be highly effective $-$ by a back action $-$ in causing decoherence of the very system which is responsible for the perturbation, in our case the qubit. Specifically, Eqs. (\ref{qubitRho}) and (\ref{Loschmidt}) show that the qubit gets maximally entangled with the environment when ${\cal L}(t) \rightarrow 0$,
with a complete loss of coherence. More formally, consider the purity of the qubit, $P(t) = Tr_q(\rho_q^2(t))$. A straightforward calculation reveals that
$P(t)=1-2|c_{g}c_{e}|^2[1-{\cal L}(t)]$, from which one again reads off the imprint of the Loschmidt echo on the qubit decoherence. It is worth pointing out that by choosing the initial environmental state $|\phi_{\text{env}}(0)\rangle$ as an eigenstate of the unperturbed Hamiltonian $H_{\text{env}}$, the LE in Eq. (\ref{Loschmidt}) codifies a quantum quench \cite{Gogolin2016} with the perturbed Hamiltonian $H_{\text{env}}^{(\delta)}$ as quench Hamiltonian.

In the following we shall identify the conditions under which the decay of the Loschmidt echo of the QCC is at its largest, favoring a fast decoherence of the qubit state with a resultant ``quantum-to-classical" transition.

\section{Loschmidt echo of the quantum compass chain}

\subsection{Preliminaries}

To calculate the LE we should first diagonalize the environment Hamiltonian.
For this purpose we use the Jordan-Wigner transformation
\bea
\no
\sigma^{+}_{n}&\!=\!&(\sigma^{x}_{n}\!+\!{\it i}\sigma^{y}_{n})/2\!=\!\prod_{m=1}^{n-1}\left(-\sigma^{z}_{m}\right)c_{n}^{\dagger},\\
\label{JW}
\sigma^{-}_{n}&\!=\!&(\sigma^{x}_{n}\!-\!{\it i}\sigma^{y}_{n})/2\!=\!\prod_{m=1}^{n-1}c_{n}\left(-\sigma^{z}_{m}\right), \\
\no
\sigma^{z}_{n}&\!=\!&2c_{n}^{\dagger}c_{n}\!-\!1,
\eea
to map the Hamiltonian of the QCC, $H_{\text{env}}$ in (\ref{QCChamiltonian}), onto a free fermion model \cite{You2014a}.
The transformation is exact, allowing us to write
\begin{multline}
H_{\text{env}}\!=\!\sum^{N/2}_{n=1}
\big(J_{o}c^{\dagger}_{2n}c_{2n-1}\!+\!J_{e} c^{\dagger}_{2n+1}c_{2n}\!+\!J_{o} e^{-i\theta} c^{\dagger}_{2n-1}c^{\dagger}_{2n}\\
+J_{e} e^{i\theta} c^{\dagger}_{2n}c^{\dagger}_{2n+1}+h(c^{\dagger}_{2n-1}c_{2n-1}+ c^{\dagger}_{2n}c_{2n}) \!+\! \mbox{H.c.}\big).
\label{eq3}
\end{multline}

Next we partition the fermionic chain into diatomic unit cells (labelled by $n=0,1,2,....N/2$) and introduce two independent fermions at each cell,
$c_{n}^{A}\equiv c_{2n}$ and $c_{n}^{B}\equiv c_{2n+1}$.
Inserting these operators and their adjoints into Eq. (\ref{eq3}) and Fourier transforming, one obtains
\begin{multline}
\label{eq5}
H_{\text{env}}=\sum_{k}\big[J_k(\theta) c_{k}^{A\dag}c_{-k}^{B\dag}+L_k c_{k}^{A\dag}c_{k}^{B} \\
+h(c_{k}^{A\dag}c_{k}^{A}+c_{k}^{B\dag}c_{k}^{B}) + \mbox{H.c.}\big],
\end{multline}
where $J_k(\theta)\equiv J_{o}e^{\oi\theta}-J_{e}e^{\oi(k-\theta)}$, $L_k \equiv J_{o}+J_{e}e^{\oi k}$, with $0 \le \theta \le \pi$. Having imposed periodic boundary conditions on the original QCC Hamiltonian
in (\ref{QCChamiltonian}), its fermionic counterpart in (\ref{eq5}) is defined with antiperiodic boundary conditions, $c^{A/B}_{N+1}\!=\!- c^{A/B}_1$, for which
$k=\pm 2\pi(2n+1)/N$ with $n=0,1,... N/4-1$. To simplify notation, from now on we suppress the dependence on the parameter $\theta$.
%
%

By bringing in the Nambu spinor $\Gamma^{\dagger}~=~
(c^{A\dagger}_{k}\, c^{B\dagger}_{k}\, c^{A}_{-k}\, c^{B}_{-k})$,
$H_{\text{env}}$ in (\ref{eq5}) can be expressed on Bogoliubov-de Gennes form
as ${\small H_{\text{env}}=\sum_{k>0}\Gamma^{\dagger}H(k)\Gamma}$, with
%
%
%
%

%
\bea
\label{BdG}
H(k)=
 \left(
  \begin{array}{cccc}
    -2h & L_{k} & 0 & J_{k} \\
    L_{k}^{\ast} &  -2h & -J_{-k} & 0 \\
    0 & -J_{-k}^{\ast} & 2h & -L^{\ast}_{-k} \\
    J_{k}^{\ast} & 0 & -L_{-k} & 2h \\
  \end{array}
\right),
\eea

The Bloch matrix $H(k)$ is easily diagonalized, yielding the quasiparticle-form of the QCC Hamiltonian,
$H_{\text{env}}=\sum_{\alpha=1}^{4}\sum_{k}\varepsilon^{(\alpha)}_{k}\gamma_{k}^{(\alpha) \dag}\gamma_{k}^{(\alpha)}$, where $\gamma_{k}^{(\alpha) \dag}$ and $\gamma_{k}^{(\alpha)}$
are linear combinations of the electron operators in the Nambu spinor with respective energy dispersions
$\varepsilon^{(1)}_{k}\!\!=\!\!-\varepsilon^{(4)}_{k}\!\!=\!\!-\sqrt{a_k+\sqrt{a_k^{2}-b_k}}$ and $\varepsilon^{(2)}_{k}\!\!=\!\!-\varepsilon^{(3)}_{k}\!\!=\!\!-\sqrt{a_k-\sqrt{a_k^{2}-b_k}}$,
where $a_k\!=\! 8h^{2}+|J_{k}|^{2}+|L_{k}|^{2}+|J_{-k}|^{2}+|L_{-k}|^{2}$ and $b_k=4\Big[(4h^{2}-|L_{k}|^{2})^{2}+4h^{2}(J_{k}^{2}+J_{-k}^{2})+J_{k}^{2}J_{-k}^{2}-J_{k}^{\ast}J_{-k}L_{k}^{2}-J_{k}J_{-k}^{\ast}L_{-k}^{2}\Big]$. Note that the Bloch matrix $H^{(\delta)}(k)$ (and the corresponding quasiparticle dispersions) of the perturbed QCC Hamiltonian $H_{\text{env}}^{(\delta)}$ is obtained from $H(k)$ by simply replacing $h$ by $h+\delta$.

The QCC ground state $|\psi_0\rangle$ is realized by filling up the negative-energy quasiparticle states, $|\psi_0\rangle = \prod_k \gamma_k^{(1) \dag} \gamma_k^{(2) \dag} |0\rangle$, where $|0\rangle$ is the Bogoliubov vacuum annihilated by the $\gamma_k$:s \cite{Jafari2016}. While excited states can be similarly obtained, their construction becomes quite cumbersome within the Bogoliubov-de Gennes formalism. An alternative approach was pioneered by Sun \cite{Sun2009}. One here takes off from the observation that the QCC Hamiltonian can be written as a sum of commuting Hamiltonians $H_k$, obtained by grouping together terms in (\ref{eq5}) with opposite signs of $k$,
\begin{multline}
\label{commH}
H_k=J_{k}c_{k}^{A\dag}c_{-k}^{B\dag}+L_{k}c_{k}^{A\dag}c_{k}^{B} +J_{-k}c_{-k}^{A\dag}c_{k}^{B\dag}\\ +L_{-k} c_{-k}^{A\dag}c_{-k}^{B}
+ h(c_k^{q\dag}c_k^{q\phantom\dagger} + c_{-k}^{q\dag}c_{-k}^{q\phantom\dagger}) +\mbox{H.c},
\end{multline}
with $q=A,B$ summed over. In the same way as above, the effective potential from the qubit can be included by simply replacing $h$ by $h+\delta$ in (\ref{commH}).
Since $H_k$ conserves the number parity (even or odd number of electrons), it is sufficient to consider the even-parity
subspace of the Hilbert space, spanned by
{\small
\begin{align}
\no
|\varphi_{1,k}\rangle&=|0\rangle,\!&\!|\varphi_{2,k}\rangle&=c_{k}^{A\dag}c_{-k}^{A\dag}|0\rangle,~|\varphi_{3,k}\rangle=c_{k}^{A\dag}c_{-k}^{B\dag}|0\rangle,\\
\no
|\varphi_{4,k}\rangle&=c_{-k}^{A\dag}c_{k}^{B\dag}|0\rangle,~\!&\!|\varphi_{5,k}\rangle&=c_{k}^{A\dag}c_{-k}^{B\dag}|0\rangle, ~|\varphi_{6,k}\rangle=c_{k}^{A\dag}c_{k}^{B\dag}|0\rangle,\\
\label{eq7}
|\varphi_{7,k}\rangle&=c_{-k}^{A\dag}c_{-k}^{B\dag}|0\rangle,~\!&\!|\varphi_{8,k}\rangle&=c_{k}^{A\dag}c_{-k}^{A\dag}c_{k}^{B\dag}c_{-k}^{B\dag}|0\rangle.
\end{align}}
Given this basis, the eigenstates $|\psi_{m,k}\rangle$ of $H_k$ can be written as $|\psi_{m,k}\rangle=\sum_{j=1}^{8}v_{m,k}^{(j)}|\varphi_{j,k}\rangle,$
%
%
%
\begin{figure*}[t!]
\centerline{\includegraphics[width=0.34\textwidth]{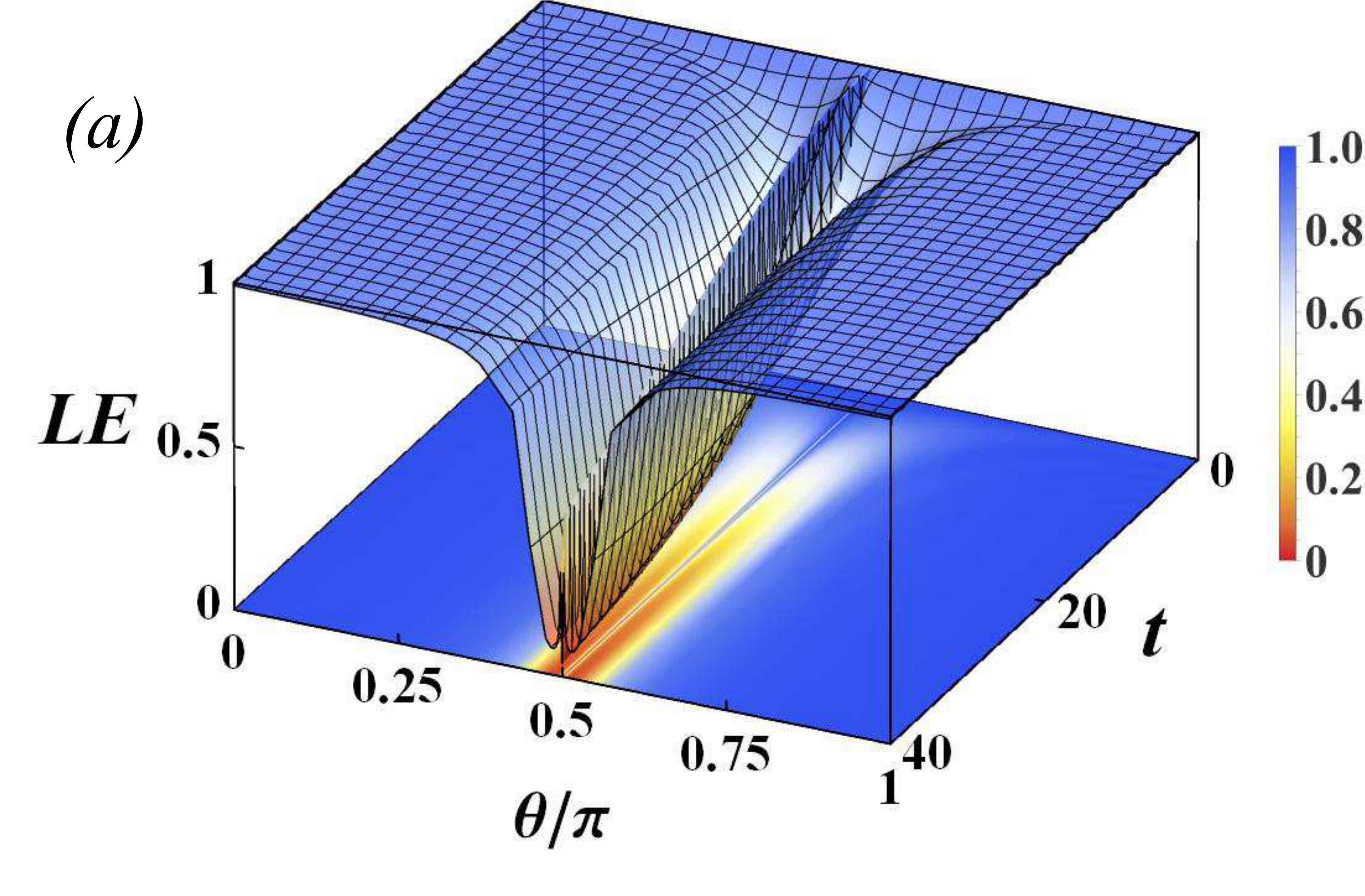}
\includegraphics[width=0.32\textwidth]{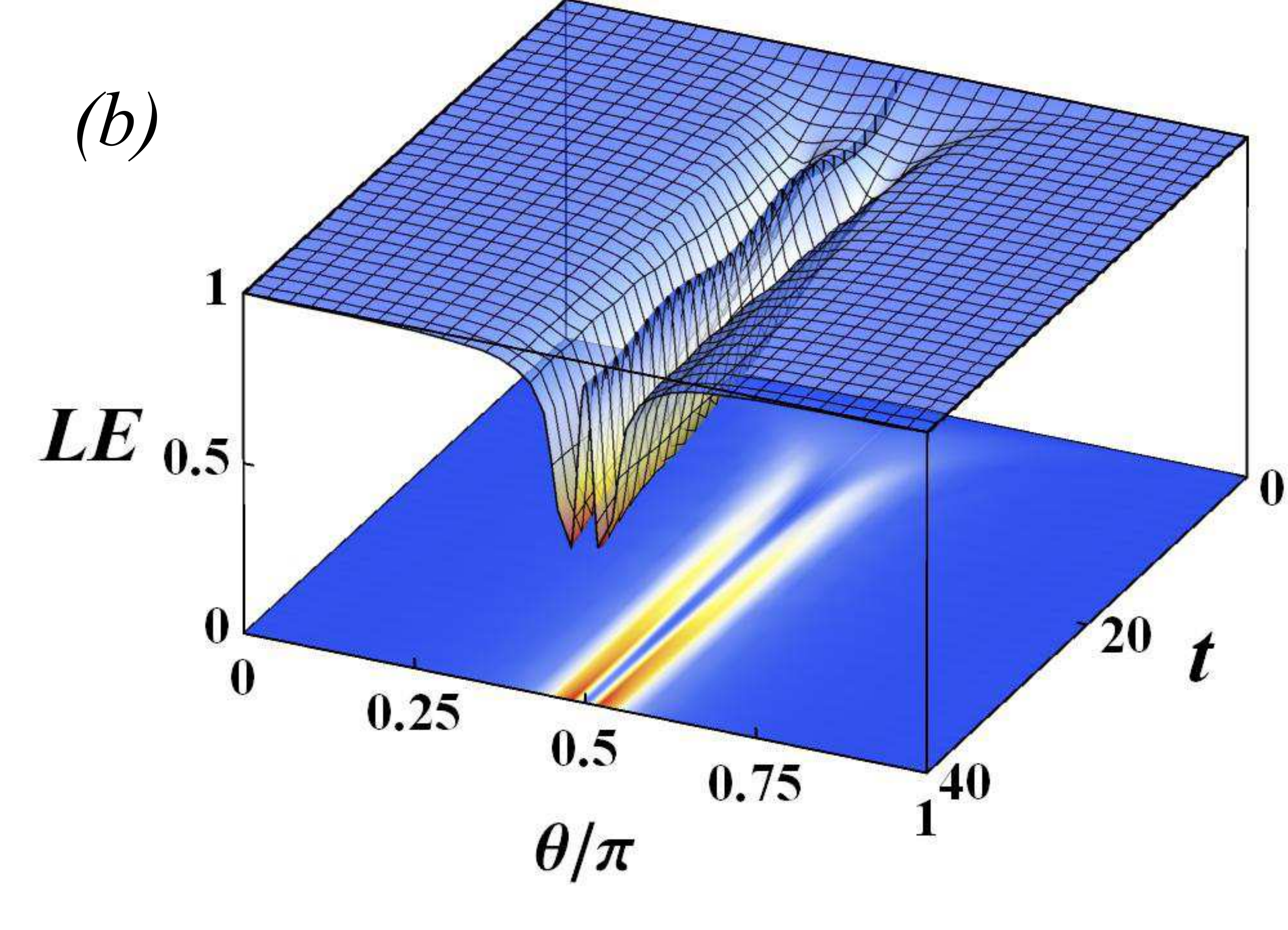}
\includegraphics[width=0.32\textwidth]{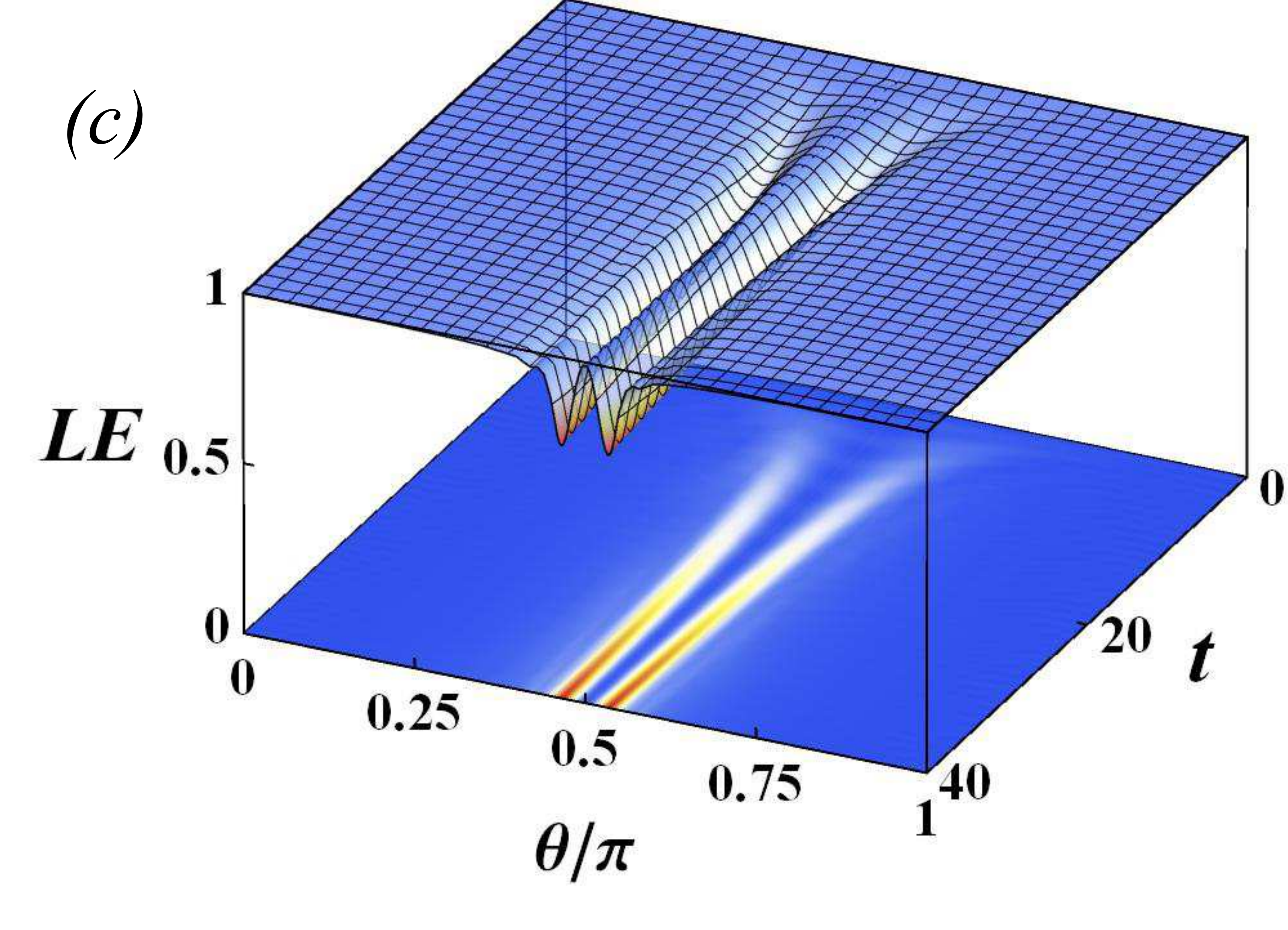}}
\caption{ (Color online) Three-dimensional plots of
the LE in Eq. (\ref{Loschmidt}) as a function of time $t$ and spin-component mixing angle $\theta$,
with $J_{o}\!=\!1$, $h=0$, $\delta=0.01$, $N\!=\!400$, and (a) $J_{e}\!=\!1$,
(b) $J_{e}\!=\!1.2$, (c) $J_{e}\!=\!2$.}
\label{fig1}
\end{figure*}
%
where $v_{m,k}^{(j)}$ with $j=1,... 8$ and $m=0,... 7$ are functions of $J_e,J_o, h, \theta$, and $k$. A straightforward calculation reveals that there is a four-fold degenerate zero-energy level below (above) which there are two bands with negative (positive) energies $\epsilon_{0,k}=(-\epsilon_{7,k})=\varepsilon^{(1)}_{k}+\varepsilon^{(2)}_{k}$ and $\epsilon_{1,k}=(-\epsilon_{6,k})=\varepsilon^{(1)}_{k}-\varepsilon^{(2)}_{k}$ respectively, with $\varepsilon_k^{(1)}$ and $\varepsilon_k^{(2)}$ the quasiparticle energies defined after Eq. (\ref{BdG}).

Before plunging ahead with the calculation of the LE $-$ equipped with the results derived above $-$ let us briefly review some pertinent facts about the QCC model. In the absence of a magnetic field, the model enjoys a $\mathbb{Z}_2$ symmetry when $\theta=\theta_c=\pi/2$ for which the model is critical \cite{You2014a}. Here a quantum phase transition (QPT) takes place between two gapped spin-liquid phases $-$ each characterized by large short-range spin correlations in the $x$- and $y$-direction, respectively $-$ for parameter values where the
model exhibits maximum frustration of interactions.
On the critical line $\theta_c=\pi/2$ in $(\theta,J_e/J_o)$-space, the ground state has a macroscopic degeneracy of $2^{N/4}$ when $J_{o}\neq J_{e}$, which gets enlarged to $2\times2^{N/4}$ at the isotropic point (IP) $J_{o}=J_{e}$. Away from the IP a gap of size $|J_{e}-J_{o}|$ opens at the zone boundaries $k= \pm \pi$, explaining the lower degeneracy in this case.
Adding a magnetic field $h$, the massive degeneracy of the ground state collapses to a two-fold degeneracy at the critical values $h_{c}=\pm \cos(\theta)\sqrt{J_{o}J_{e}}$ \cite{You2014a}.

\subsection{Loschmidt echo: zero magnetic field}

To obtain an expression for the LE ${\cal L}(t)$ in (\ref{Loschmidt}) that is practical for computations, we first make a mode decomposition of the QCC ground state $|\phi_{\text{env}}\rangle$,
\bea
|\phi_{\text{env}}\rangle = \prod_{0\le k \le \pi} |\psi_{0,k}\rangle,
\eea
using that $|\psi_{0,k}\rangle$ is the lowest-energy eigenstate of $H_k$ in (\ref{commH}).
Introducing a notation for the LE and the eigenstates of $H_k$ where the presence (or absence) of magnetic field $h$ and/or perturbing potential $\sim \delta$ is made explicit, ${\cal L}(h+\delta,t)$ and $|\psi_{m,k}(h+\delta)\rangle$ respectively, the LE in (\ref{Loschmidt}) can then be decomposed as
\bea
\label{eq8}
{\cal L}(\delta,t)&=&\prod_{0\le k \le \pi} {\cal L}_k(\delta,t),\\
\no
{\cal L}_k(\delta,t)&=&\Big|\frac{1}{N_{0,k}(0)} \langle\psi_{0,k}(0)|
e^{-iH_{\text{env}}^{(\delta)}t} |\psi_{0,k}(0)\rangle\Big|^{2}\\
\no
&=&\Big|\frac{1}{N_{0,k}(0)}
\sum_{m=0}^{7}\frac{e^{-i \epsilon^{(\delta)}_{m,k}t}}{N_{m,k}(\delta)}|
\langle\psi_{m,k}(\delta)|\psi_{0,k}(0)\rangle|^{2}\Big|^{2}
\label{eq8b} \\
\eea
where $N_{m,k}(h+\delta)=(\sum_{j=1}^{8}|v_{m,k}^{j}(h+\delta)|^{2})^{1/2}$
is the normalization factor of the eigenstate $|\psi_{m,k}(h+\delta)\rangle$, and $\epsilon^{(\delta)}_{m,k}$ is the eigenvalue of $H_k$ (with $h$ replaced by $h+\delta$ in (\ref{commH}))
corresponding to $|\psi_{m,k}(\delta)\rangle$.

A straightforward calculation shows that ${\cal L}_k(\delta,t)$ can be expressed as
\bea
\label{eq9}
{\cal L}_k(\delta,t)&&=
|1-A_{0,k}\sin^{2}[(\varepsilon_{k}^{(1)}(\delta)+\varepsilon_{k}^{(2)}(\delta))t]\\
\no
&&-B_{0,k}\sin^{2}[(\varepsilon_{k}^{(1)}(\delta)+\varepsilon_{k}^{(2)}(\delta))t/2]\\
\no
&&-A_{1,k}\sin^{2}[(\varepsilon_{k}^{(1)}(\delta)-\varepsilon_{k}^{(2)}(\delta))t]\\
\no
&&-B_{1,k}\sin^{2}[(\varepsilon_{k}^{(1)}(\delta)-\varepsilon_{k}^{(2)}(\delta))t/2]\\
\no
&&-C_{k}\sin^{2}[\varepsilon_{k}^{(2)}(\delta)t]-D_{k}\sin^{2}[\varepsilon_{k}^{(1)}(\delta)t]|.
\eea
Here $A_{0,k}, B_{0,k}, A_{1,k}, B_{1,k}, C_k$ and $D_k$ are products of linear combinations of the state overlaps $F_{m,k}=|\langle\psi_{m,k}(\delta)|\psi_{0,k}(0)\rangle|^{2}$ $(m=0,...,7)$
(for details, see the Appendix), with $\varepsilon_k^{(\alpha)}(\delta) \,(\alpha=1,2)$ being energies of the quasiparticles filling up the ground state of $H_{\text{env}}^{(\delta)}$.
The second filled quasiparticle band in the ground state becomes dispersionless along the critical line $\theta_c= \pi/2$, with $\epsilon_k^{(2)}=0$.
For this case the LE reduces to the simple form
\bea
\no
{\cal L}_k(\delta,t)=|1-A_{c,k}\sin^{2}(\varepsilon_{k}^{(1)}(\delta)t)-B_{c,k}\sin^{2}(\frac{\varepsilon_{k}^{(1)}(\delta)t}{2})|,\\
\label{eq10}
\eea
with $A_{c,k}=A_{0,k}+A_{1,k}+D_{k}$ and $B_{c,k}=B_{0,k}+B_{1,k}$.
%
%

Having obtained explicit formulas for the LE in (\ref{eq9}) and (\ref{eq10}), we are now ready to numerically probe some representative cases. Choosing $\delta=0.01$, the behavior of ${\cal L}(\delta,t)$ versus
$t$ and $\theta/\pi$ at the {\em isotropic point (IP)} {\boldmath $J_e\!=\!J_o$} is displayed in FIG. 1(a) for a chain with 400 unit cells.
It is seen from the figure that the LE decays fast at the critical point $\theta_{c}=\pi/2$ of the {\it unperturbed} QCC Hamiltonian.
However, the decay of the LE is even faster slightly off the unperturbed critical point, where the LE exhibits two subvalleys. An analysis reveals
that the extra valleys occur at the critical points $\theta_c = \arccos(\pm \delta/\sqrt{J_eJ_o})$ of the {\it perturbed} QCC Hamiltonian $H^{(\delta)}_{\text{env}}$.
Thus, the criticality of the perturbed and the unperturbed environmental Hamiltonians is here the common feature
linked to an accelerated decay of the LE. While many numerical studies suggest that criticality of an environment enhances the decay of the
LE \cite{Quan2006,Yuan2007a,Cucchietti2007,Sun:2007aa,Ma2007,Mostame2007,Rossini2008,Haikka2012,Sharma2012}, to the best of our knowledge our result is the first that explicitly displays an enhanced decay both at the unperturbed {\it and} perturbed critical point  of a model environment.

Let us now study the LE {\em away from the IP} {\boldmath $(J_{e}\neq J_{o})$}. The case $J_o=1.0, J_e=1.2$, with (as before) $\delta = 0.01$ and $N=400$, is plotted in FIG. 1(b).
Similar to the case of the IP, the decay of the LE is again seen to be at its maximum at the critical points of the perturbed Hamiltonian ($\theta_{c}=\arccos(\pm \delta/\sqrt{J_{e}J_{o}})$).
Different from the IP, however, the decay of the LE at the critical point of the unperturbed Hamiltonian ($\theta_c=\pi/2$) shows no enhancement but fluctuates around a constant value close to unity. {\em This challenges the common notion that a critical environment always leads to a fast decay of the LE \cite{Quan2006}, and by that, a fast decoherence of the system that couples to it \cite{Yuan2007a}.}
Notably, increasing the length of the chain or the observation time does not change this conclusion.

What is the reason for this unexpected result? To find out, let us first go back to Eqs. (\ref{eq8}), (\ref{eq8b}), and (\ref{eq9}) and try to understand, from a mathematical point of view, how these equations control the decay of the LE.

\begin{figure*}[t]
\centerline{\includegraphics[width=0.37\linewidth]{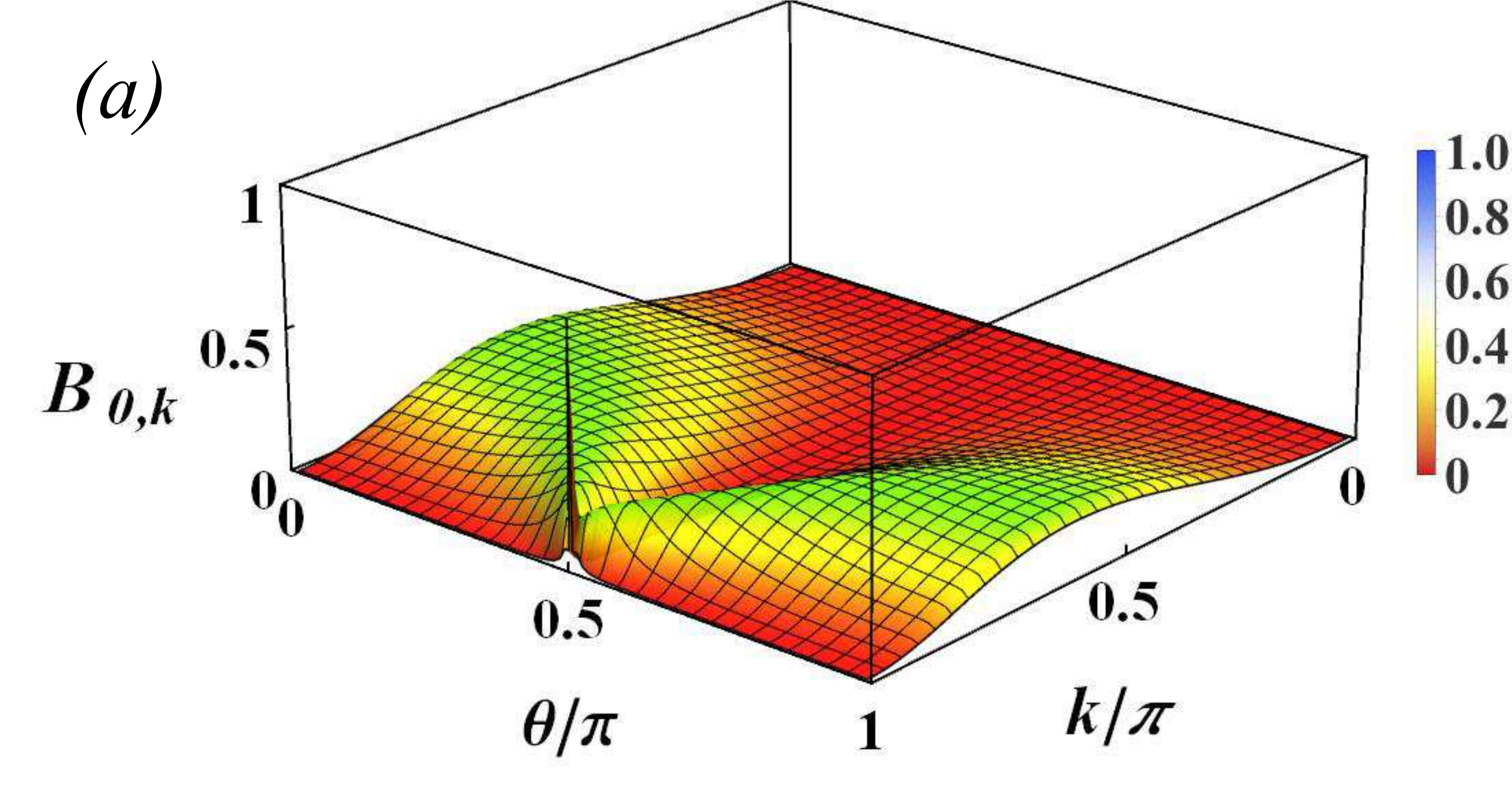}
\includegraphics[width=0.34\linewidth]{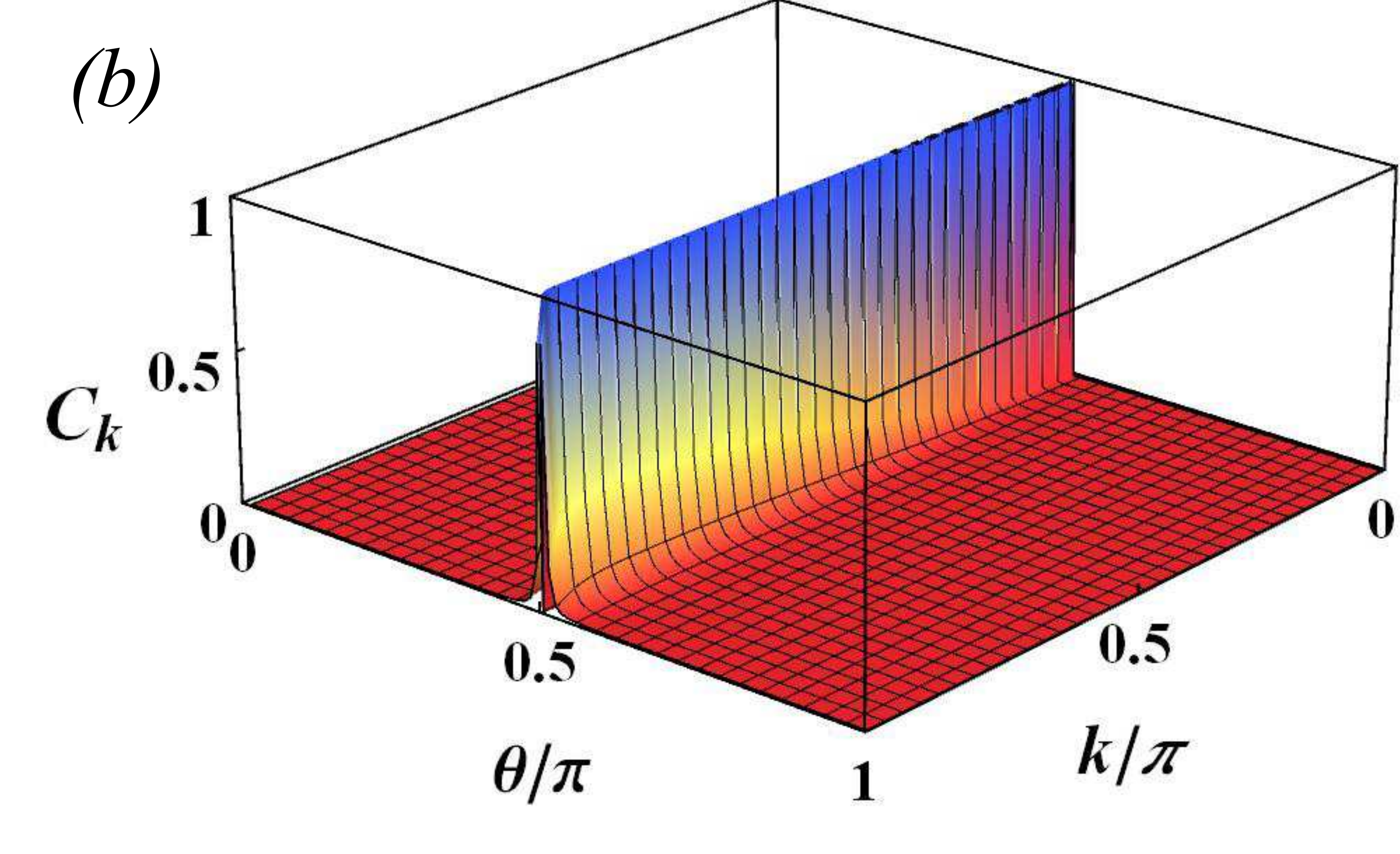}
\vspace{0.4cm}
\includegraphics[width=0.28\linewidth]{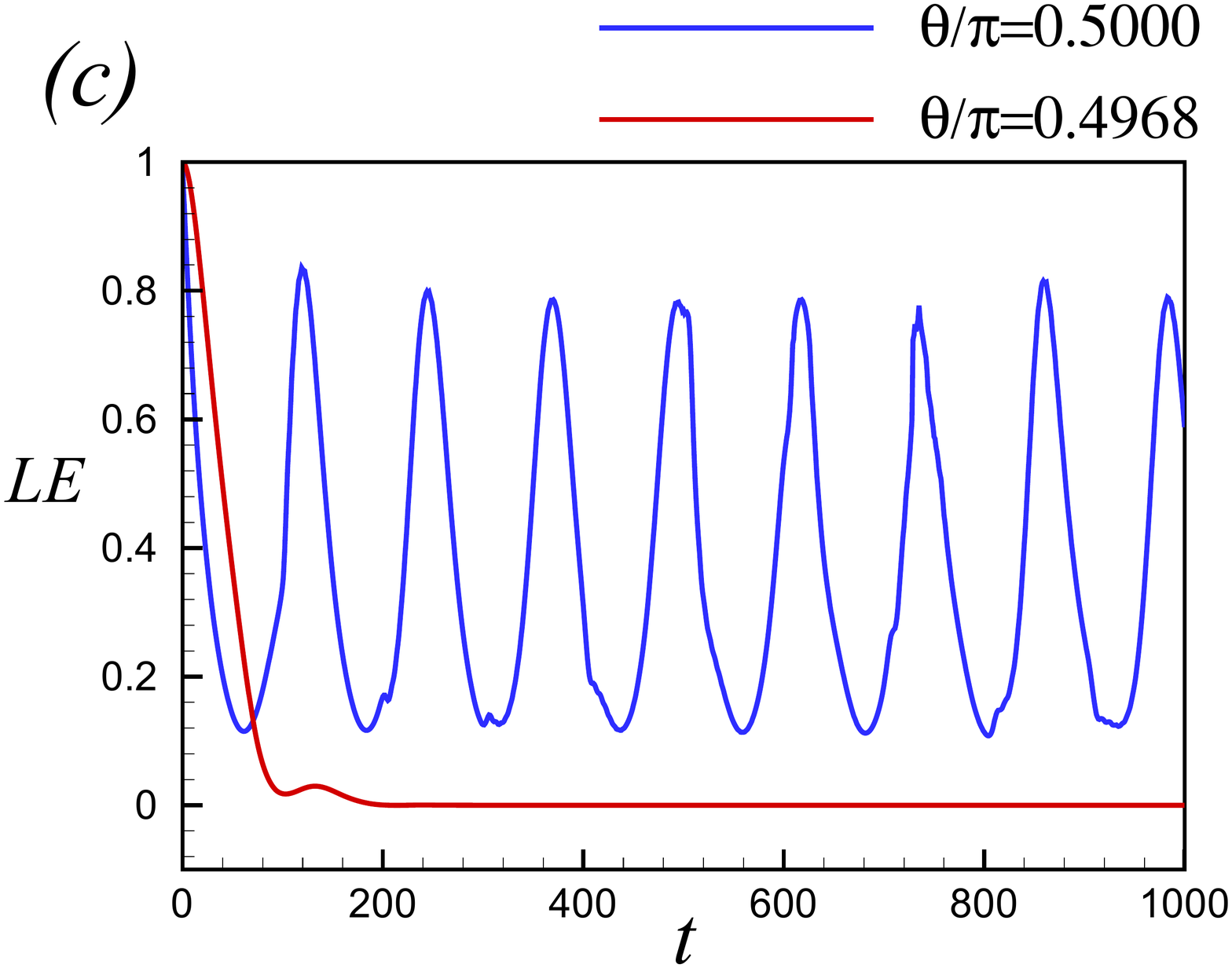}}
\caption{ (Color online) Oscillation amplitude (a) $B_{0,k}$ and (b) $C_{k}$
in the mode decomposition, Eq. (\ref{eq9}), of the LE \\ as function of crystal momentum $k$ and spin-component
mixing angle $\theta$ {\em at the isotropic point}
$J_{o}\!=\!J_{e}\!=\!1$, with qubit- environment coupling $\delta\!=\!0.01$, and with $h=0, N\!=\!400$.
(c) Time evolution of the LE, Eq. (\ref{Loschmidt}), for the same set of parameter values at
$\theta= 0.5000\,\pi$ (critical point of the unperturbed QCC) and $\theta=0.4968 \,\pi$ (critical point of the perturbed QCC).}
\label{fig2}
\end{figure*}

Since the maximum value of any $k$-mode ${\cal L}_k(\delta,t)$ is unity, it is clear from Eq. (\ref{eq8}) that it is sufficient that only a few of the modes take on very small values in order for the LE to get suppressed. As manifest in Eq. (\ref{eq9}), the actual contribution from a given $k$-mode to the LE is controlled by its oscillation terms, with a small/large value of an oscillation term implying a large/small contribution. An analysis reveals that all oscillation amplitudes $A_{0,k}$, $B_{0,k}$, $A_{1,k}$, $B_{1,k}$, $C_{k}$, and $D_{k}$, are small at the {\em IP critical point} {\boldmath $\theta_c=\pi/2$}, except for $B_{0,k}$ when approaching one of the Brillouin zone boundaries $k=\pm \pi$ at which $B_{0,k}$ reaches a sharp maximum (FIG. \ref{fig2}(a)). It follows from Eq. (\ref{eq10}) that the corresponding modes in the immediate neighborhood of a zone boundary will contribute constructively/destructively over time intervals where $\sin^{2}(\varepsilon_{k}^{(1)}(\delta)t)$ is small/large. Thus, by the periodicity of the sine-function, the LE is expected to exhibit periodic revivals, signaling a non-Markovian reduced dynamics of the qubit
with a backflow of information from the environment \cite{Haikka2012}. This expectation is well confirmed numerically, cf. FIG. \ref{fig2}(c).
%
%
\begin{figure*}
\centerline{\includegraphics[width=0.34\linewidth]{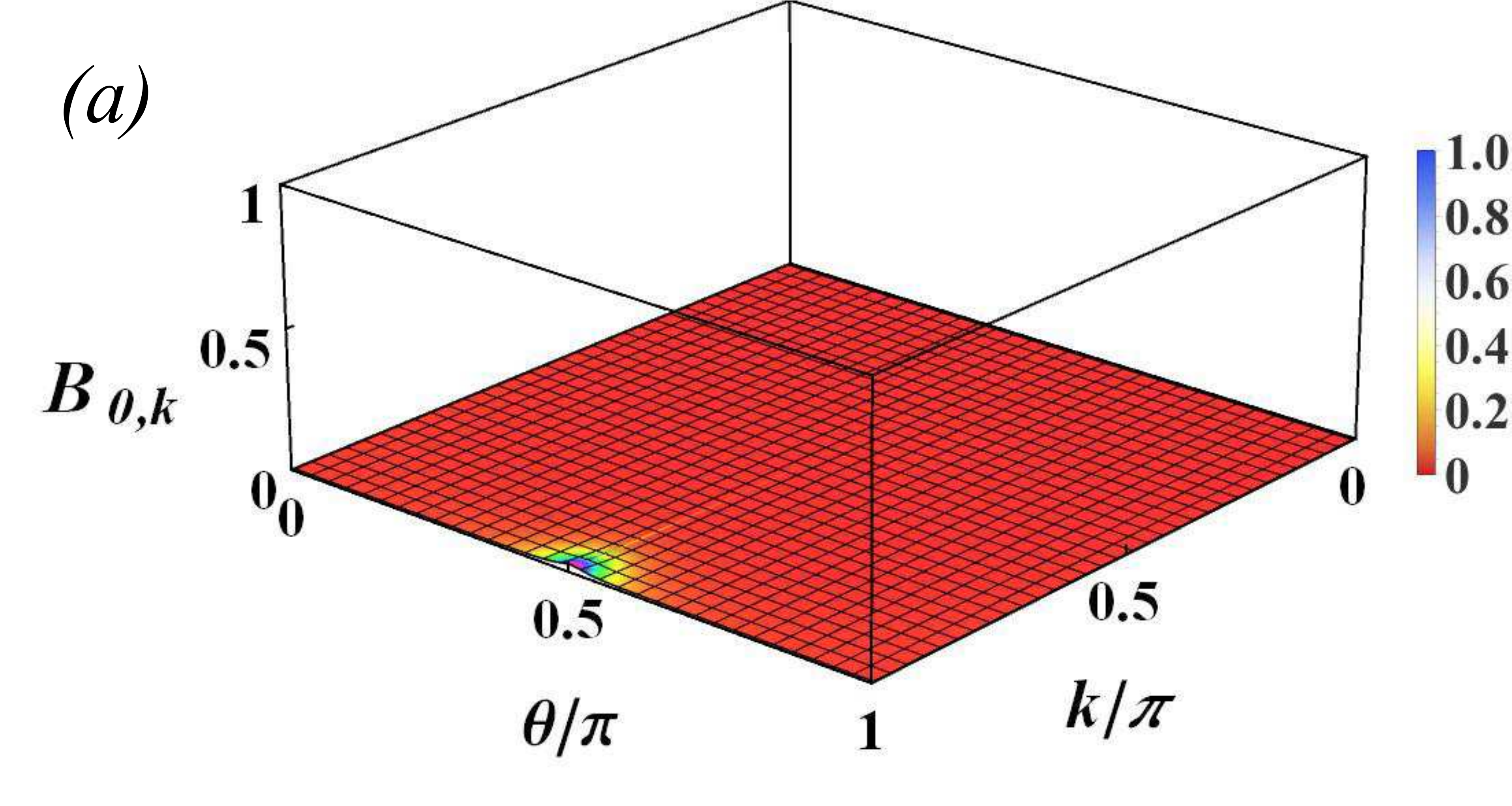}
\includegraphics[width=0.32\linewidth]{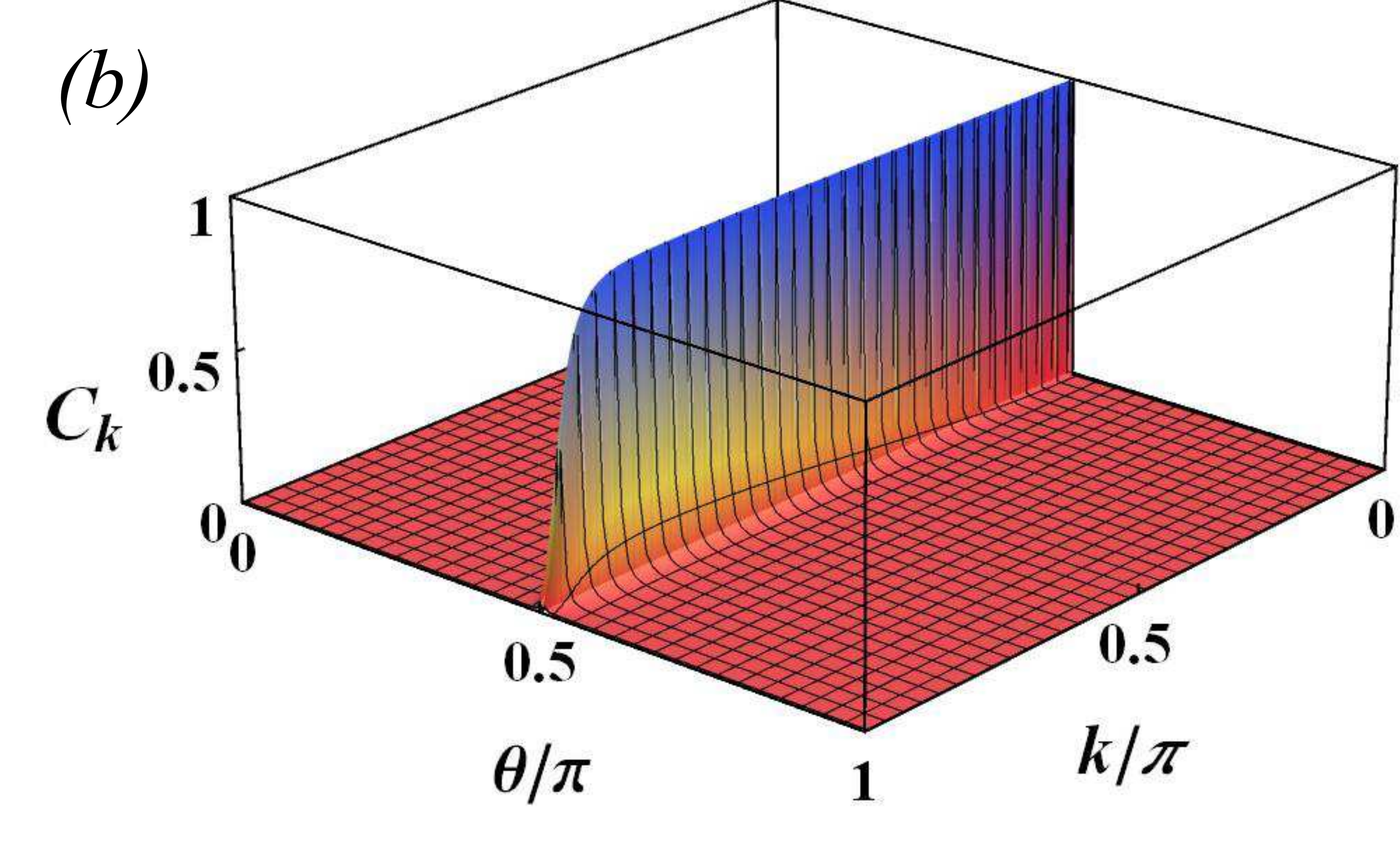}
\includegraphics[width=0.28\linewidth]{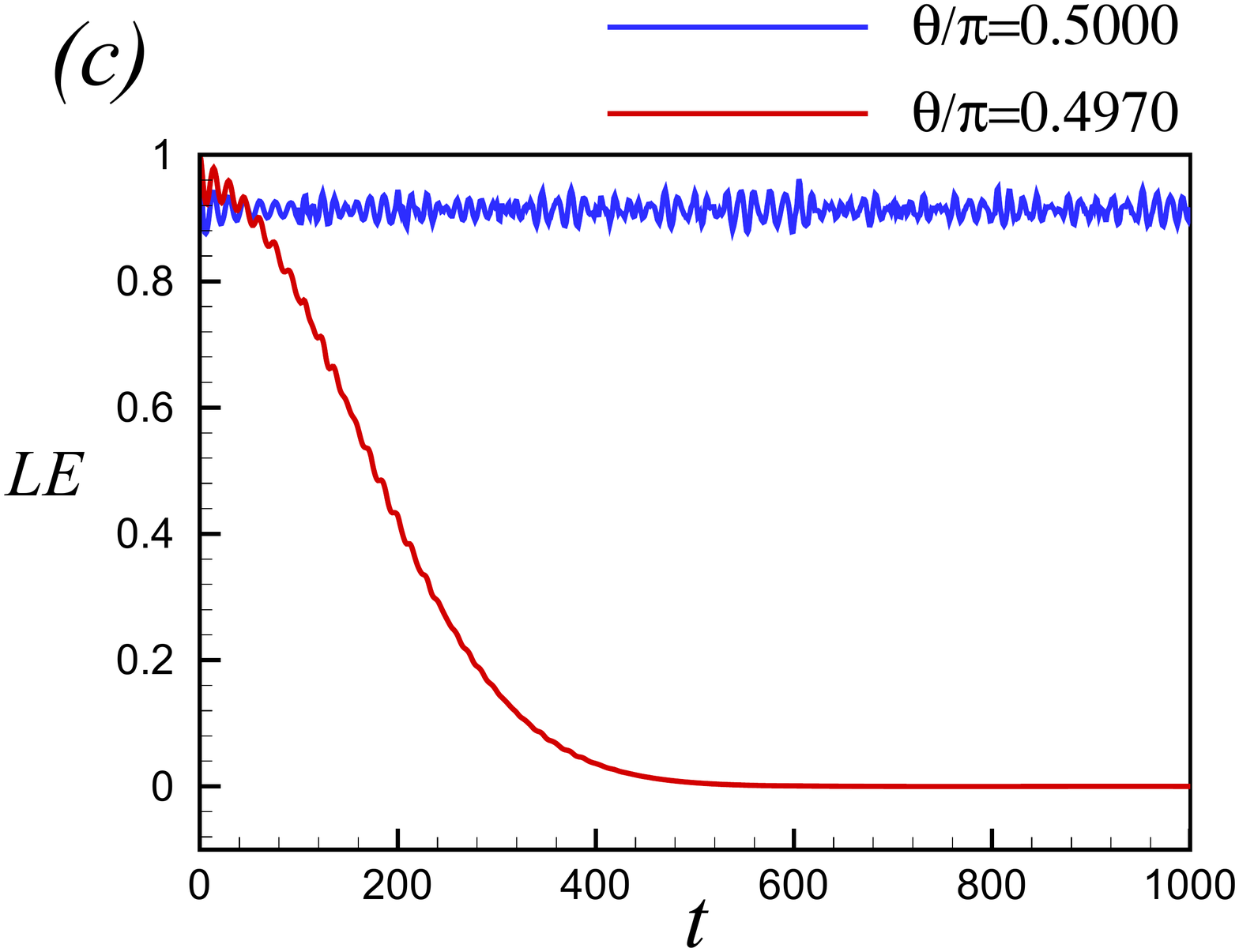}}
\caption{ (Color online) Oscillation amplitude (a) $B_{0,k}$ and (b) $C_{k}$
in the mode decomposition, Eq. (\ref{eq9}), of the LE as function of crystal momentum $k$ and spin-component
mixing angle $\theta$ {\em away from the isotropic point}, $J_o\!=\!1$ and $J_e\!=\!1.2$,
with qubit-environment coupling $\delta=0.01$, and with $h=0, N=400$.
(c) Time evolution of the LE for the same set of parameter values for
$\theta \!=\! 0.5000\,\pi$ (critical point of the unperturbed QCC) and $\theta_{c}\!=\!0.4970\, \pi$ (critical point of the
perturbed QCC).}
\label{fig3}
\end{figure*}
%
\begin{figure}
\includegraphics[width=0.66\linewidth]{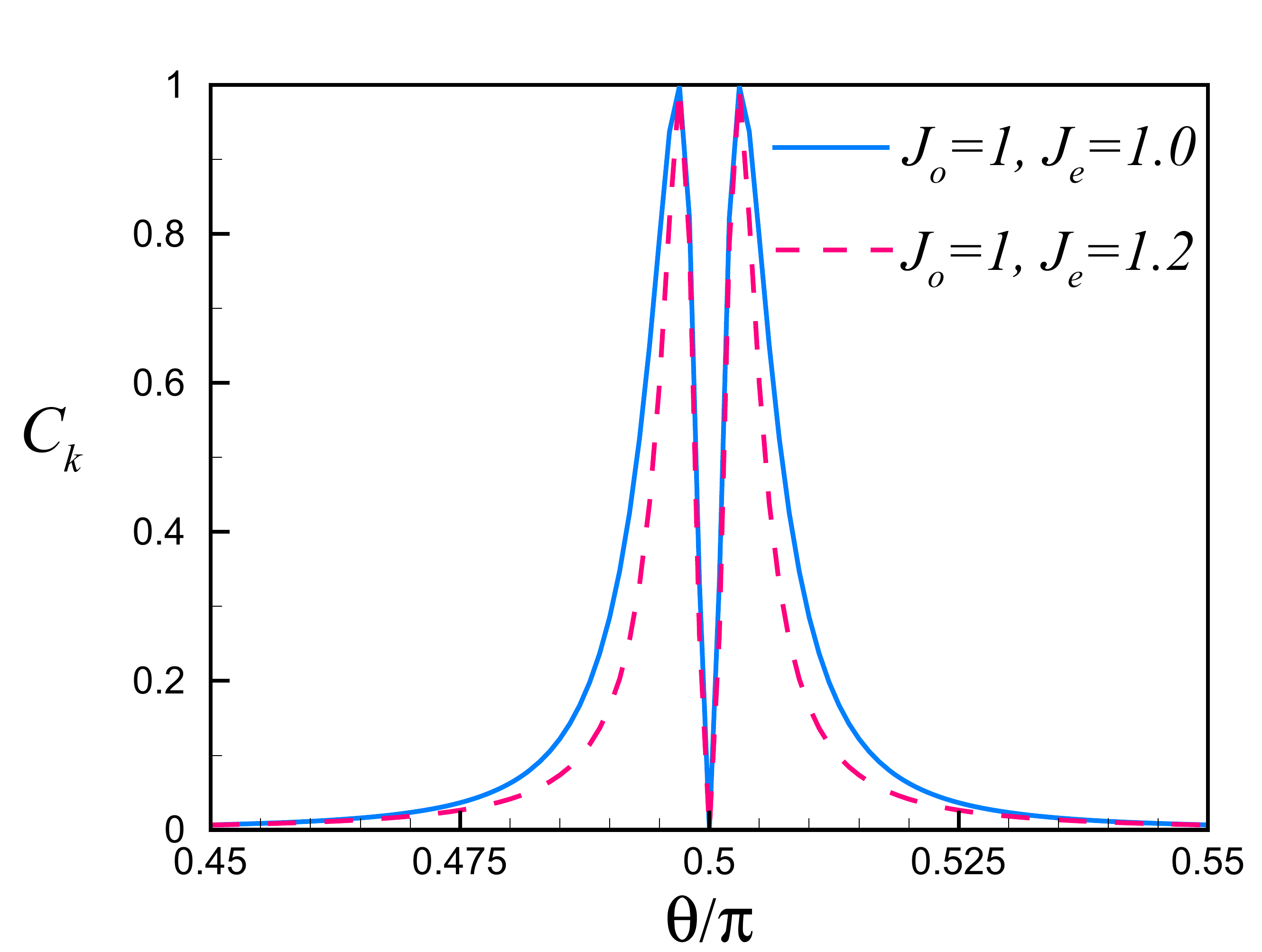}
\caption{ (Color online) Cross sections of FIGs. \ref{fig2}(b) (blue) and \ref{fig3}(b) (red) showing $C_k$ versus $\theta$ at $k=0$.}
\label{fig8}
\end{figure}
%
It is actually instructive to unearth the revival period from Eq. (\ref{eq9}), explaining why FIG. 1(a) suggests a monotonic decay of the LE for $\theta_c=\pi/2$ at the IP, while, in fact, as revealed by the blue graph in FIG. 2(c), it exhibits a stable and distinct revival structure when going to larger time scales. Following Ref. [\onlinecite{Jafari2017}], we make the Ansatz $\varepsilon_{k=\pi}^{1}(\theta_{c})\,t/2=m\pi$, with $m$ an integer and with $k=\pi$ the mode with the largest oscillation amplitude ($B_{0,k}$ at the BZ boundary). Taylor-expanding $\varepsilon_{\pi-p\delta k}^{1}(\theta_{c})
\approx \varepsilon_{\pi}^{1}(\theta_{c})-\partial_k \epsilon_{k}^{1}(\theta_{c})|_{\pi}\, p\delta k$,
one realizes that $B_{0,k}$-terms of nearby $k$-modes are strongly suppressed when
$t$ is a multiple of $Na/v_{g}$, with $v_{g}=\partial_k \epsilon_{k}^{1}(\theta_{c})|_{\pi}$ the group velocity of the corresponding quasiparticle and $a=1$ the size of the unit cell, implying a revival time $T_{\text{rev}}\approx N/v_{g}$. Here $p\ll N$ are integers and $\delta k = 2\pi/N$. Putting in numbers, one obtains $T_{\text{rev}}=122$ (in arbitrary units), in excellent agreement with FIG. 2(c), however not visible on the shorter time scale of FIG. 1(a).

Focusing now on the accelerated decay of the LE at the {\em IP critical points} {\boldmath $\theta_c = \arccos(\pm \delta/\sqrt{J_eJ_o})$} of the perturbed Hamiltonian $H_{\text{env}}^{(\delta)}$, (cf. the subvalleys in FIG. 1(a)
and the red graph in FIG. 2(c)), an analysis of Eq. (\ref{eq9}) shows that it is caused by the oscillation term $\sim C_k$. As illustrated in FIG. 2(b) for $\delta=0.01$ and $J_o\!=\!J_e\!=\!1.0$,
$C_{k}$  peaks to large values at $\theta_c = \arccos(\pm \delta/\sqrt{J_eJ_o})$ for all $k$. (For a cross-sectional view at $k=0$, see FIG (\ref{fig8}).) This is to be contrasted to the structure of $B_{0,k}$ away from $\theta_c=\pi/2$, being broad and shallow, cf. FIG. 2(a).
The revival time of the LE is now controlled by
the group velocity of the quasi particles which occupy the $\varepsilon_{k}^{2}$ band (corresponding to the $C_k$ amplitude, cf. Eq. (\ref{eq9})).
Since this band is almost flat at $\theta_c = \arccos(\pm \delta/\sqrt{J_eJ_o})$ with $\delta$ small and $J_eJ_o=1$ (see FIG. (\ref{fig5})(a)), the quasiparticle group velocity is exceedingly small: $v_g \sim 10^{-7}$ (in arbitrary units) for $\delta=0.01$. Considering the time scale of FIG. 2(c), the revival time which ensues, $T_{\text{rev}} \approx N/v_g \sim 10^6$, is far too large for the revivals to be picked up in this figure. Instead, the rapid decay and subsequent vanishing of the LE depicted by the red graph in this figure suggests a Markovian dynamics of the qubit. This is similar to a central spin model with the transverse field Ising chain as environment where the critical point has been found to support a purely Markovian dynamics over short initial times\cite{Haikka2012}. It is important to point out, however, that our analysis does predict that a (non-Markovian) revival structure will appear if waiting sufficiently long, signaling a backflow of information from the environment to the qubit at very large times. Admittedly, these revivals appear only on extremely large time scales at which a central spin model may no longer be a realistic model for capturing a decoherence process.
%
\begin{figure*}[t]
\centerline{\includegraphics[width=0.34\linewidth]{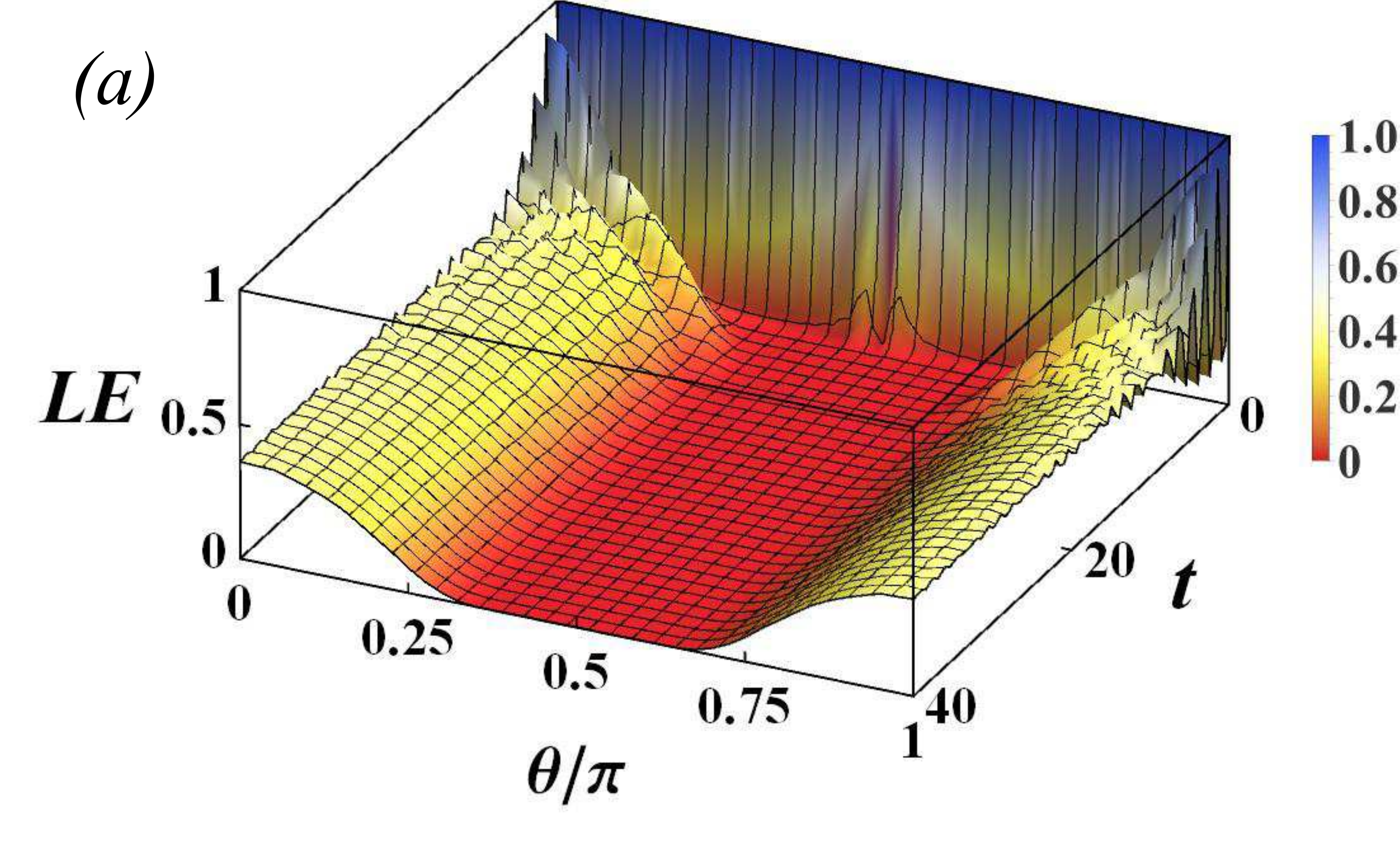}
\includegraphics[width=0.31\linewidth]{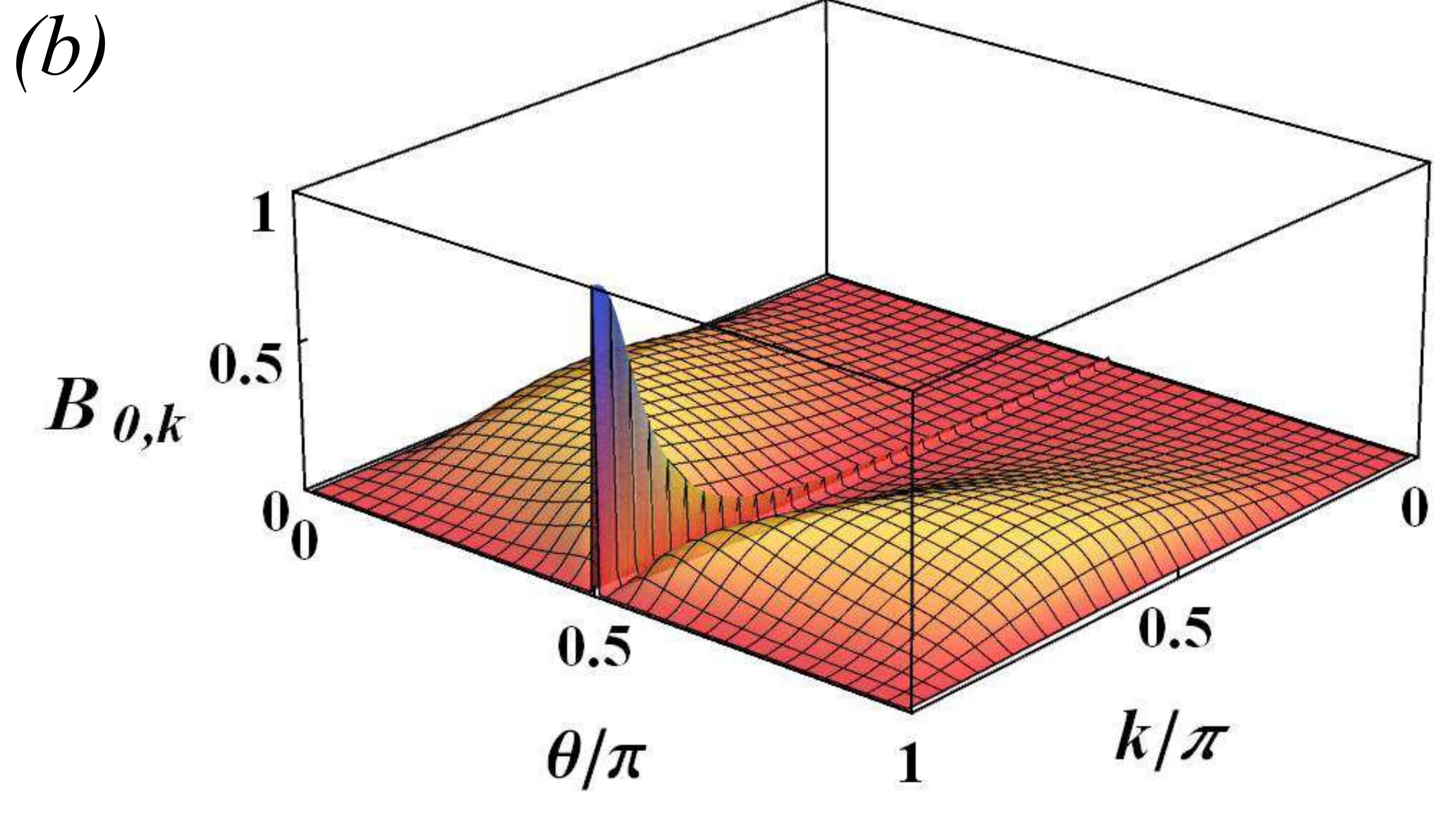}
\includegraphics[width=0.31\linewidth]{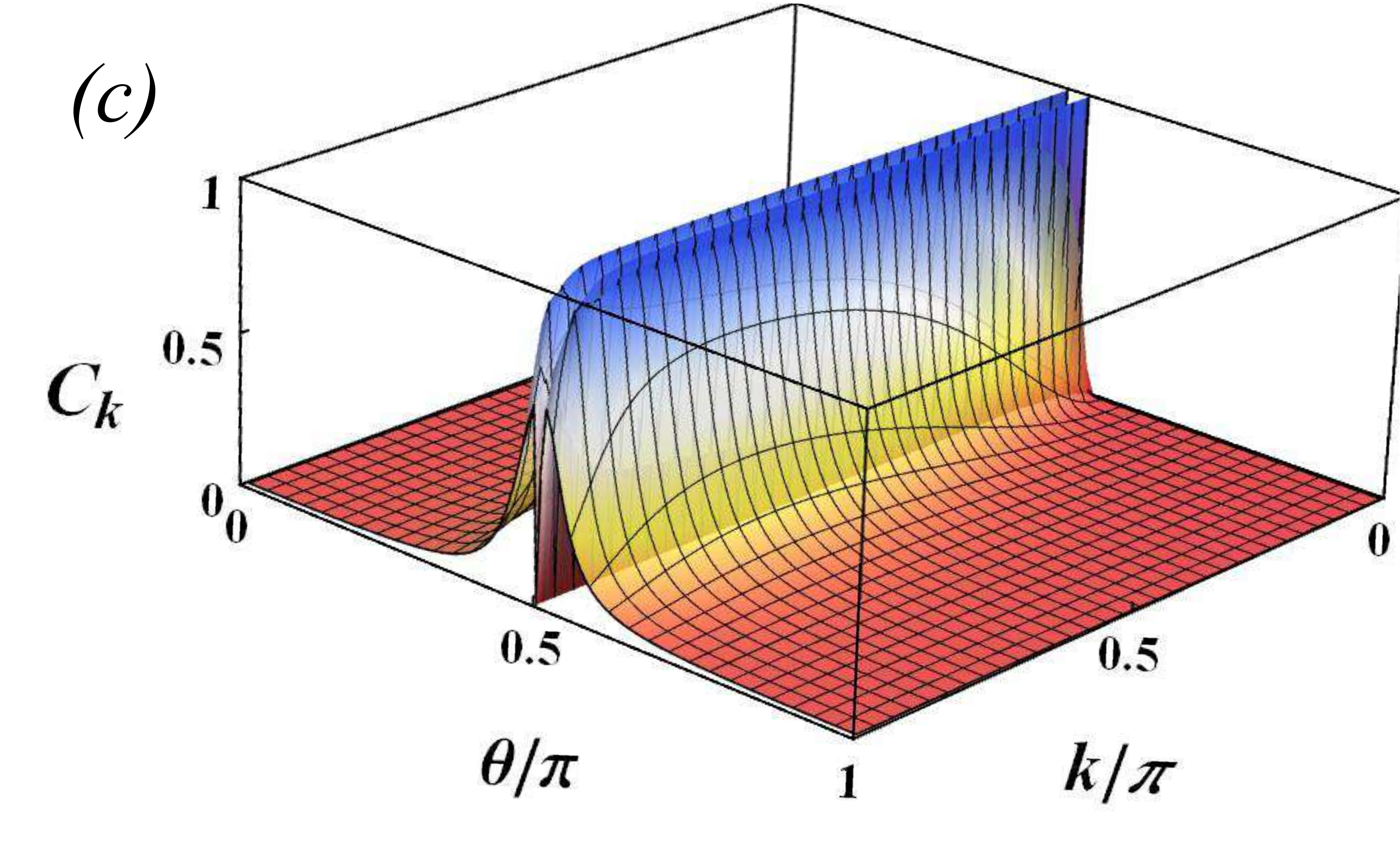}}
\caption{ (Color online) (a) Time evolution of the LE in Eq. (\ref{Loschmidt}) at the isotropic point $J_o=J_e=1$ for $\delta=0.1, h=0$, and $N\!=\!400$.
The oscillation amplitude (b) $B_{0,k}$, and (c) $C_{k}$ in the mode decomposition of the LE, Eq. (\ref{eq9}),
as a function of $k$ and $\theta$ for the same parameter values.}
\label{fig4}
\end{figure*}

Turning, finally, to the behavior of the LE {\em away from the IP}, the oscillation amplitudes $B_{0,k}$ and $C_{k}$ are plotted conundrum
in Fig. \ref{fig3}(a) and (b) for $J_{o}=1, J_{e}=1.2$.
As one can see, $B_{0,k}$ is small for all values of $\theta$ and results in the LE oscillating randomly around a
mean value close to unity at $\theta_c=\pi/2$ (Fig. \ref{fig1}(b) and (c) and Fig. \ref{fig3}(c)).
However, $C_{k}$ is still large at $\theta_c = \arccos(\pm \delta/\sqrt{J_eJ_o})$ of the perturbed theory (cf. FIG (\ref{fig8})), causing a fast decay of the LE at the critical points of the perturbed QCC Hamiltonian.
(Fig. \ref{fig3}(c)).

Before concluding this part of our discussion, let us numerically corroborate the expectation that by increasing the strength of the coupling $\delta$ between the environment and the qubit,
the decay of the LE will become faster and broader. This is strikingly illustrated for the IP in FIG. (\ref{fig4})(a), having increased $\delta$ by one order of magnitude to $\delta=0.1$.
The amplitudes of the corresponding dominating oscillation terms in the mode decomposition of the LE are depicted in Fig. \ref{fig4}(b) and (c) for all values
of $\theta$: By making the coupling $\delta$ larger the oscillation amplitudes increase and broaden, resulting in a significantly faster decay of the LE over a large parameter interval.

To understand the physics behind the different behaviors of the LE at the IP $(J_e = J_o)$ and away from the IP $(J_e \neq J_o)$, let us recall that the oscillation amplitudes
 in (\ref{eq9}) are made up of products of state overlaps $F_{m,k}=|\langle\psi_{m,k}(\delta)|\psi_{0,k}(0)\rangle|^{2}$ $(m=0,...,7)$. Knowing that $|\psi_{m,k}(0)\rangle$ is an eigenstate of $H_k$ in (\ref{commH}), implying that $\langle\psi_{m,k}(0)|\psi_{0,k}(0)\rangle = \delta_{m0}$ (up to a normalization factor), one may be tempted to argue that
$\langle\psi_{m,k}(\delta)|\psi_{0,k}(0)\rangle$ must be very small for all $m \neq 0$ since $\delta$ is a small perturbation. If this were the case, however, all oscillation amplitudes in (\ref{eq9})
would be vanishingly small for any $k$, resulting in a non-decaying LE with a value close to unity. This, as we have seen, is not the case. The argument goes wrong by the assumption that a
small perturbation can only cause a small change of a state overlap. However, a state where quasiparticles may easily be excited by a small perturbation, such as at a critical point, can dramatically
change character when perturbed and lead to sizable overlaps $\langle\psi_{m,k}(\delta)|\psi_{0,k}(0)\rangle$. Specifically, if $|\psi_{m,k}(0)\rangle$ with $m\neq 0$ is an eigenstate of the {\it unperturbed} Hamiltonian $H_k$ close to criticality (with $h=0$ in (\ref{commH})), a perturbation $|\psi_{m,k}(0)\rangle \rightarrow |\psi_{m,k}(\delta)\rangle$ may restructure the state dramatically, allowing for a finite overlap with $|\psi_{0,k}(0)\rangle$. Likewise, if $|\psi_{0,k}(\delta)\rangle$ is an eigenstate of the {\it perturbed} Hamiltonian close to its critical point (with $h=\delta$ in (\ref{commH})), $|\psi_{0,k}(0)\rangle$ may feature a very different structure with a finite overlap with $|\psi_{m,k}(\delta)\rangle$  also for $m\neq 0$. This explains why criticality of the unperturbed ($\theta_c = \pi/2$) as well as the perturbed ($\theta_c = \arccos(\pm \delta/\sqrt{J_eJ_o}))$ QCC Hamiltonian enhances the decay of the LE at the IP, making precise the expectation that the decoherence of the qubit is strongest at a critical point where the environment is most susceptible to a perturbation. Which one of the critical points that will be most effective in suppressing a LE will depend on details of the model considered, such as the particular state overlaps which enter into a given oscillation amplitude of the LE modes ${\cal L}_k(\delta,t)$. In the present case, with the QCC as environment, the decays of the LE at the IP perturbed critical points are at a maximum, followed by extremely slow revivals. Still, also the IP unperturbed critical point is quite effective in causing an initial suppression of the LE, however with fast subsequent revivals.

If we try to explain our findings {\it away from the IP} along the lines above, we are faced with an apparent conundrum. While the LE still decays at the critical point of the perturbed QCC Hamiltonian, it equilibrates around a value close to unity when at the critical point of the unperturbed theory. Why is that? Why is the critical point of the unperturbed theory now ineffective in suppressing the LE?
\begin{figure*}
\centerline{\includegraphics[width=0.33\linewidth]{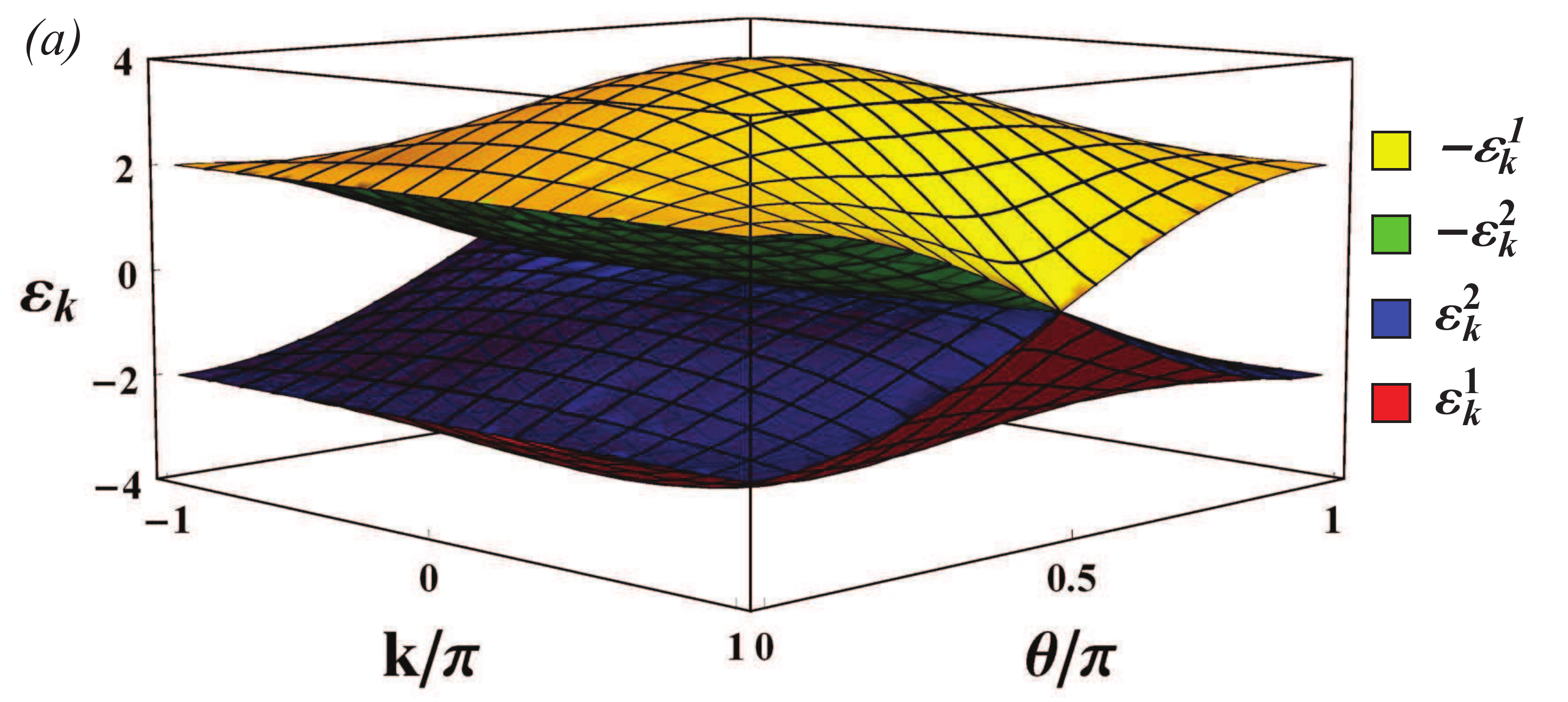}
\includegraphics[width=0.31\linewidth,]{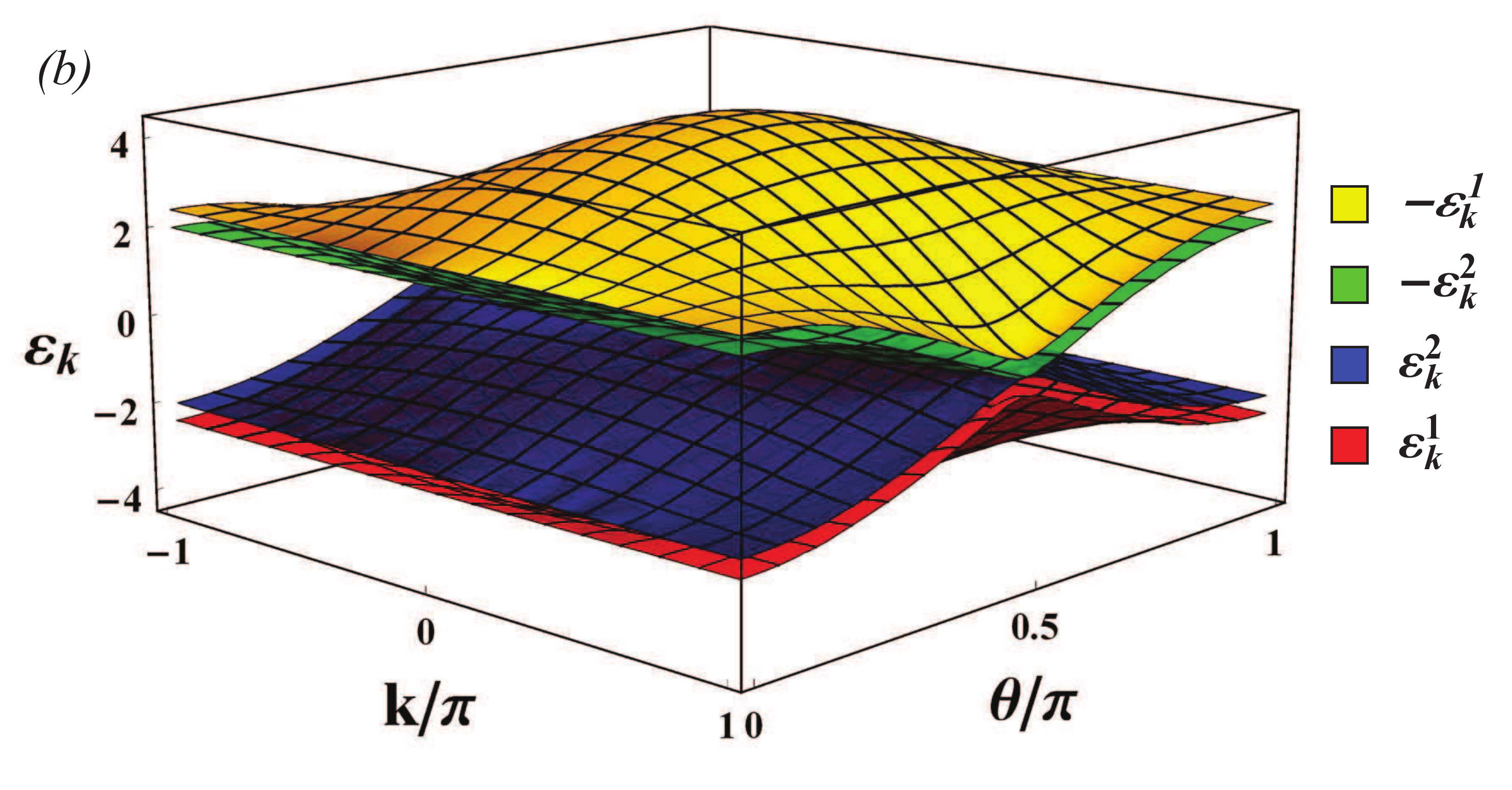}
\includegraphics[width=0.33\linewidth]{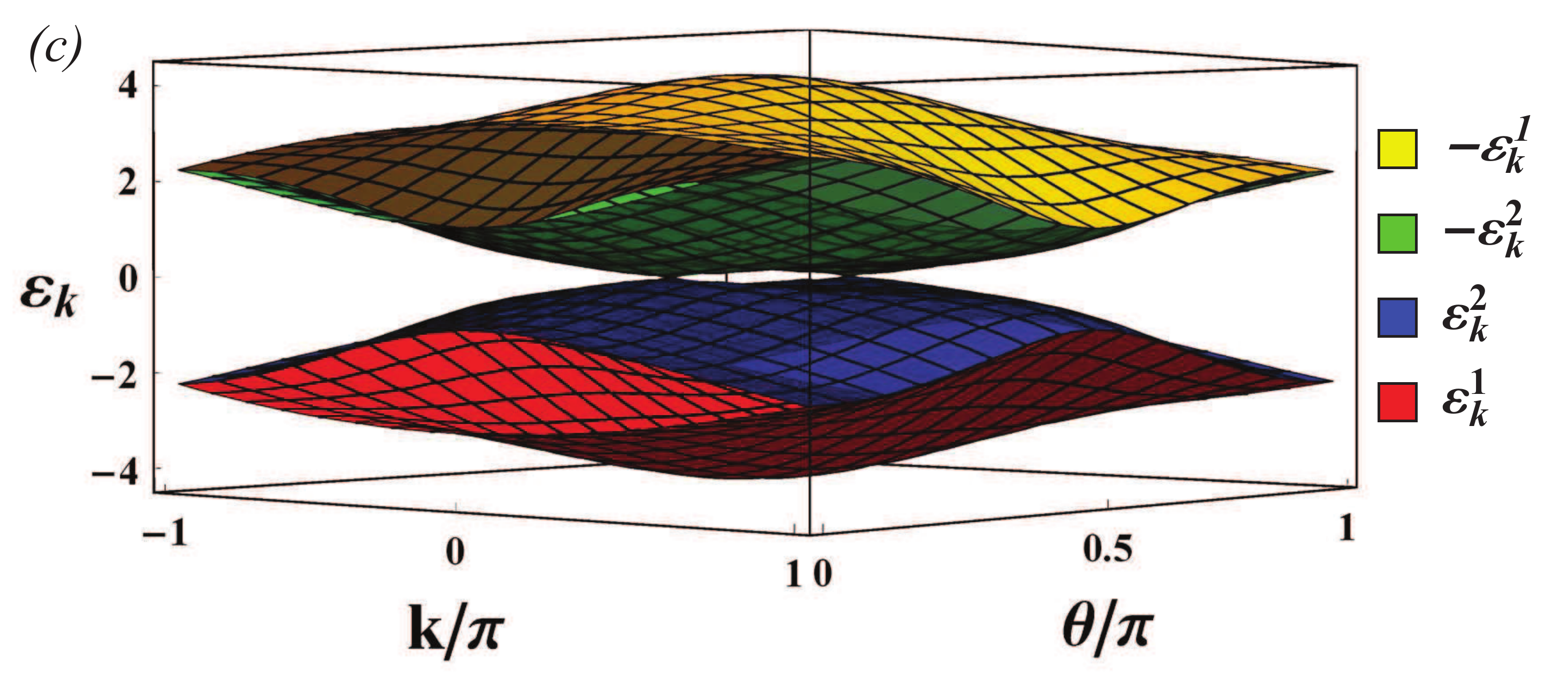}}
\caption{ (Color online) Bogoliubov-de Gennes quasiparticle spectrum $\pm\varepsilon_{k}^{1,2}(0)$
for the unperturbed QCC, Eq. (\ref{QCChamiltonian}), at (a) the isotropic point $J_{o}=J_{e}=1, h=0$,
(b) the anisotropic point $J_{o}=1, J_{e}=1.2, h=0$, and (c)
the isotropic point $J_{o}=J_{e}=1$, $h=0.5$.}
\label{fig5}
\end{figure*}
%
\begin{figure*}
\centerline{\includegraphics[width=0.23\linewidth]{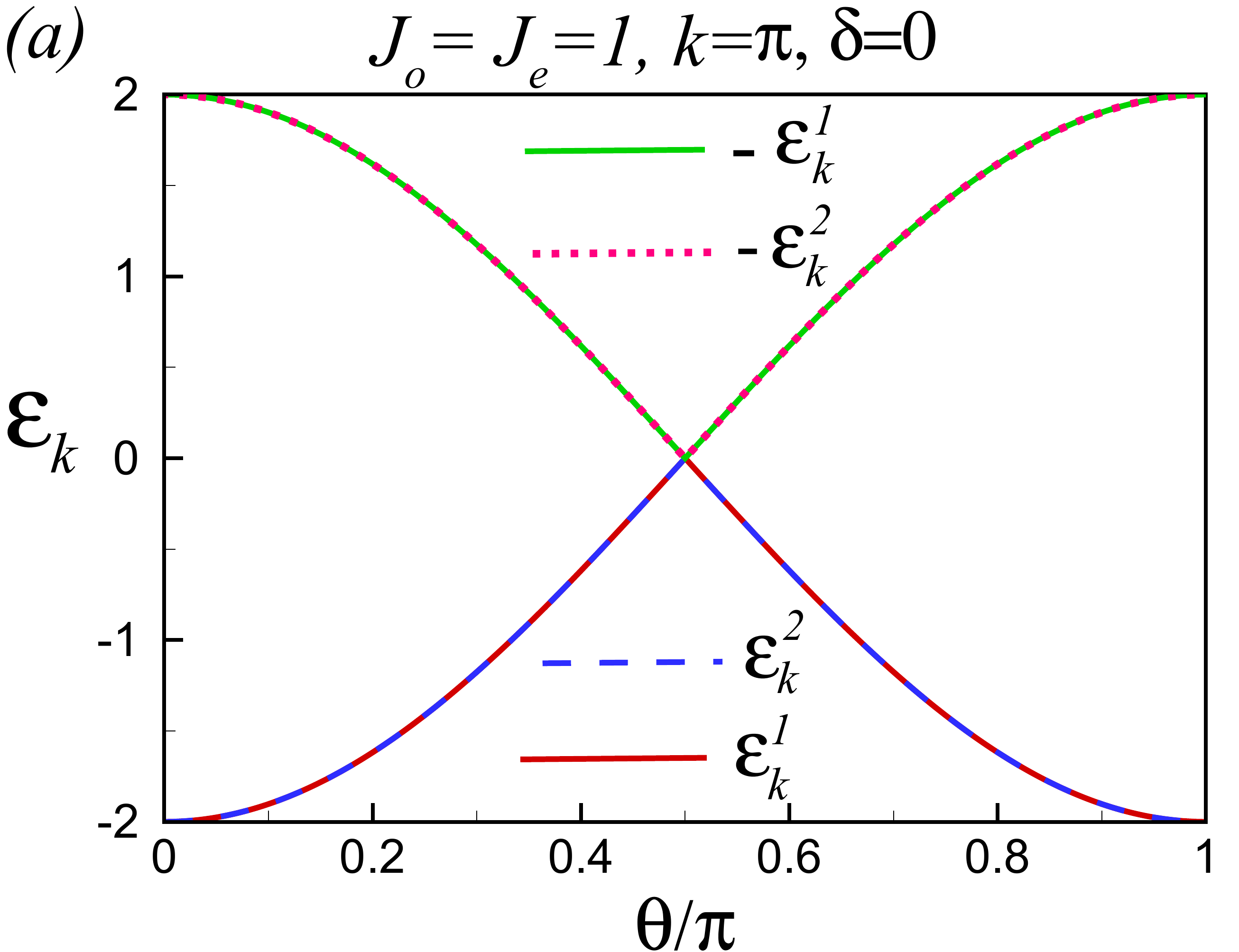}
\includegraphics[width=0.23\linewidth]{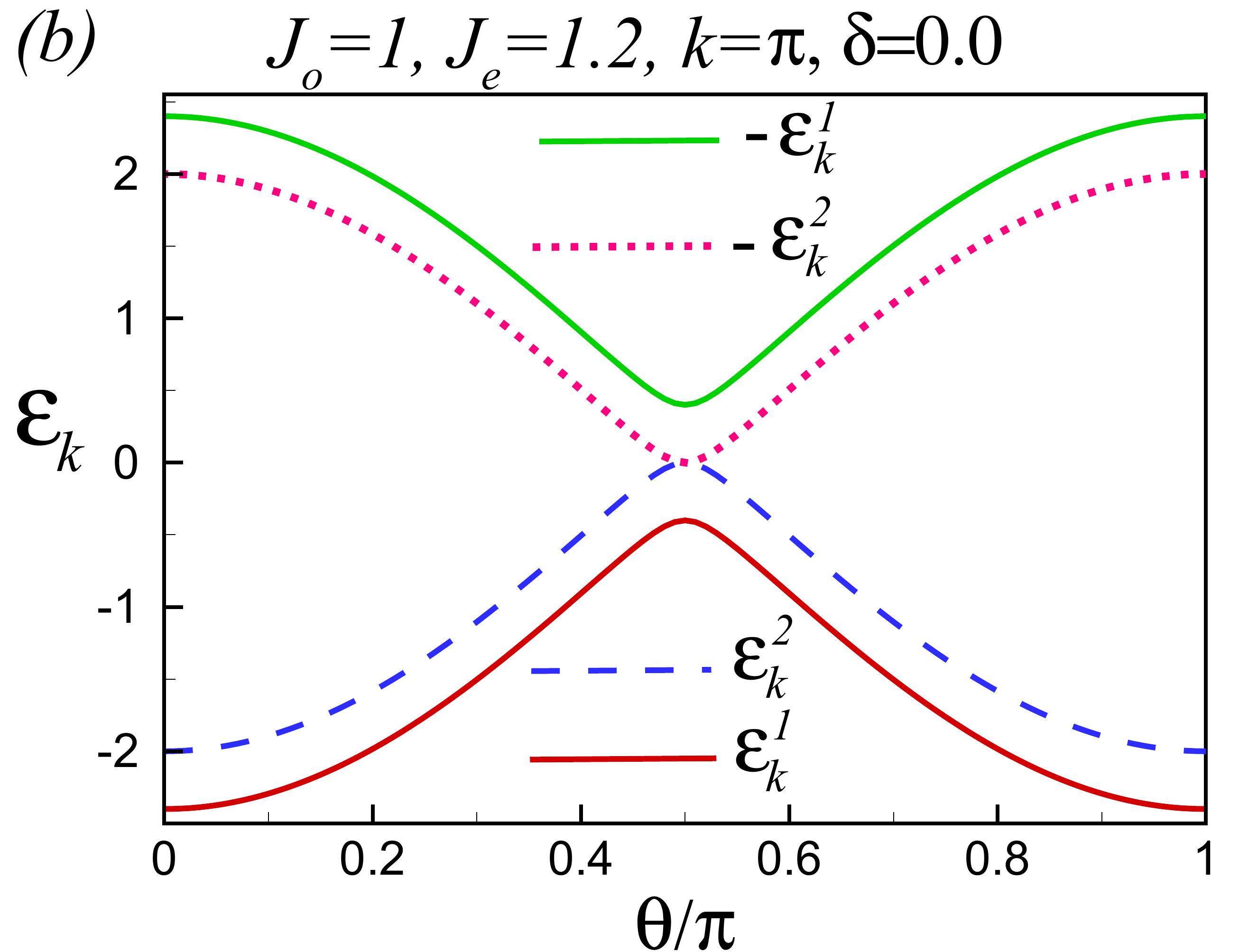}
\includegraphics[width=0.23\linewidth]{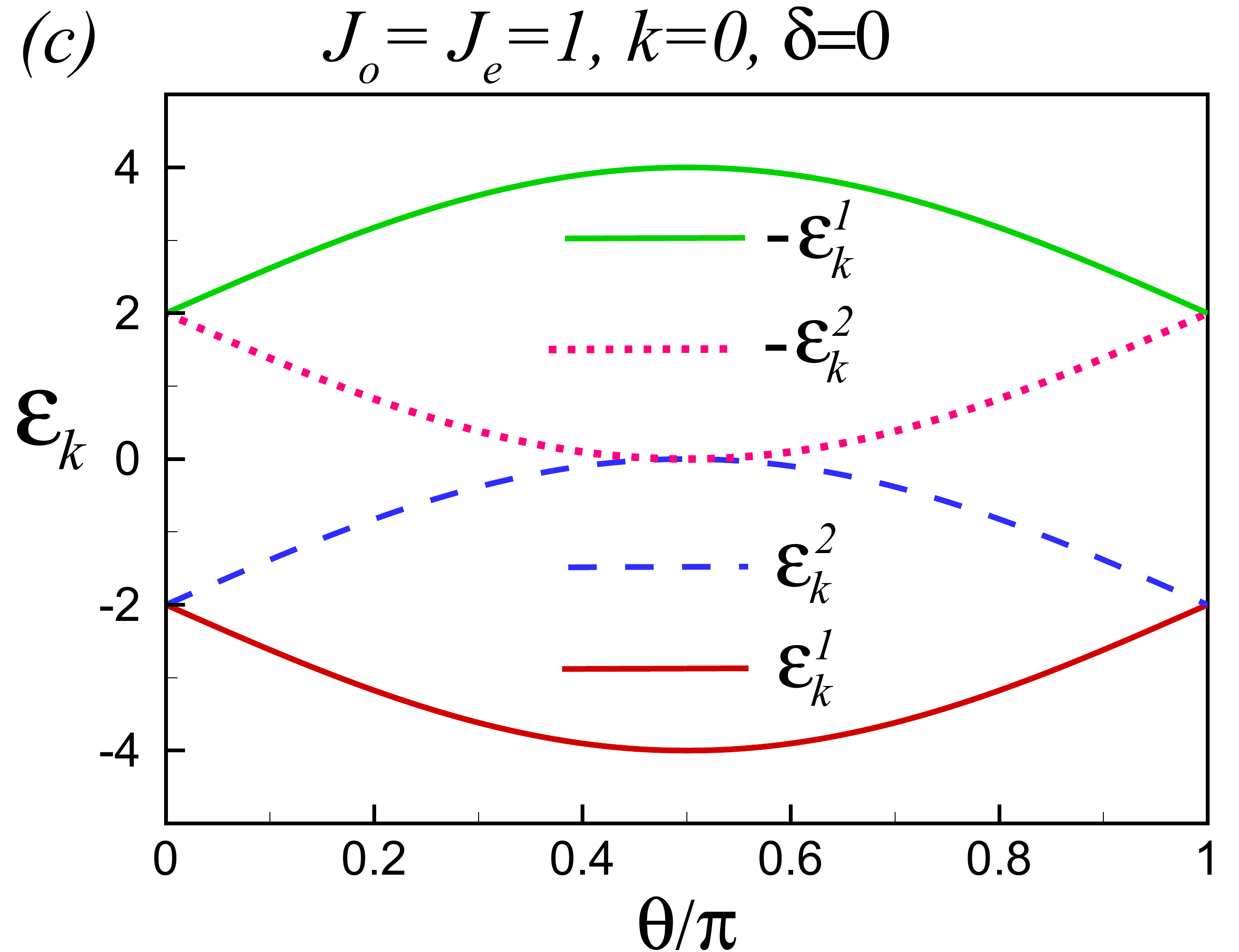}
\includegraphics[width=0.23\linewidth]{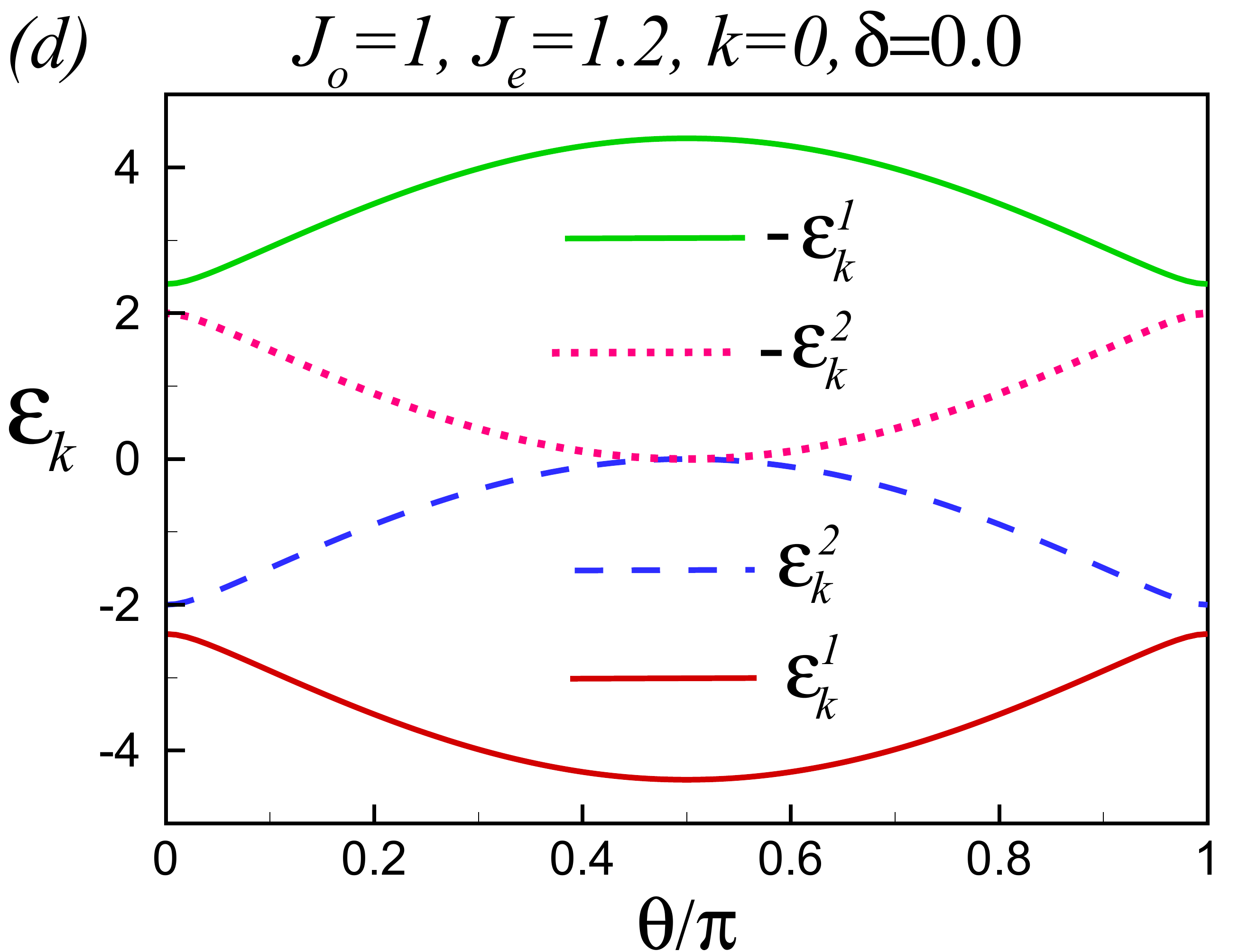}}
\caption{ (Color online) Cross section of (a) FIG. (\ref{fig5})(a) at $k=\pi$; (b) FIG. (\ref{fig5})(b) at $k=\pi$;
(c) FIG. (\ref{fig5})(a) at $k=0$; (d) FIG. (\ref{fig5})(b) at $k=0$.}
\label{fig9}
\end{figure*}

The answer can be found by inspecting the quasiparticle spectrum, FIG. (\ref{fig5}). Panel (a) shows the unperturbed QCC spectrum at the IP, where the $\varepsilon_k^{(1)}$ band (which, together with the $\epsilon_k^{(2)}$ band, is completely filled in the QCC ground state) is seen to be degenerate with the other bands at $k=\pi$ and $\theta_c=\pi/2$, thus favoring quasiparticle excitations in the neighborhood of $k=\pi$. This, as we have argued, explains why one of the IP oscillation amplitudes in the LE modes, $B_{0,k}$ as it turns out, becomes large at $k=\pi$. Now look at panel (b) of FIG. (\ref{fig5}) which displays the spectrum away from the IP, with $J_e/J_o=1.2$. Here a gap has opened up at $k=\pi$, separating the $\varepsilon_k^{(1)}$ band from that of $\varepsilon_k^{(2)}$, thus holding back quasiparticle excitations and, as a consequence, dampening the oscillation amplitudes in the LE modes. (For cross-sectional views of the spectra in FIGs (\ref{fig5})(a) and (b) at $k=\pi$, see FIGs (\ref{fig9})(a) and (b), respectively.) The filled $\varepsilon_k^{(2)}$ band is still degenerate with the next higher band for all $k$ at $\theta_c=\pi/2$. However, as evident from FIG. 3(a), the possibility of quasiparticle excitations from this band does not compensate for the loss of excitations from the $\epsilon_k^{(1)}$ band: the $B_{0,k}$ amplitude is now strongly suppressed. It is here important to note that the $\varepsilon_k^{(2)}$ band is dispersionless for all $k$ at $\theta_c=\pi/2$. Thus, the quasiparticles from this band cannot contribute significantly to the time-dependent parts of the oscillation terms at the IP for small $\delta$ and hence cannot influence the revival structure of the LE.

Different from the scenario at the critical point of the unperturbed QCC Hamiltonian, the LE at the critical point of the perturbed theory, $\theta_c=\arccos(\pm \delta/\sqrt{J_eJ_o})$, is controlled by the           $\epsilon_k^{(2)}$ band and the mode oscillation amplitude $C_k$, at the IP (FIG. 2(b)) as well as away from the IP (FIG. 3(b)). In both cases the $\epsilon_k^{(2)}$ band of the unperturbed QCC Hamiltonian is gapless at $k=0$,
(cf. FIG. (\ref{fig5})(a) and (b), respectively, with cross sections in FIG. (\ref{fig8})(c) and (d)), making quasiparticles easy to excite. For small $\delta$, when the two unperturbed critical points are close to $\pi/2$, the gap to excitations away from $k=0$ is extremely small, still allowing for an avalanche of quasiparticle excitations with a concurrent dramatic restructuring of the eigenstates. This is the reason for the almost constant and large value of the $C_k$ amplitude across the halved Brillouin zone in FIGs 2(b) and 3(b). As we have already discussed, the fact that the controlling $\varepsilon_k^{(2)}$ band is almost flat for all $k$ close to $\pi/2$ explains why the decay of the LE at the critical points of the perturbed QCC Hamiltonian appears to be monotonic: the group velocity $v_g$ of the quasiparticles is very small, resulting in exceedingly large revival periods, also for small finite systems.
%
\begin{figure*}
\centerline{\includegraphics[width=0.34\linewidth]{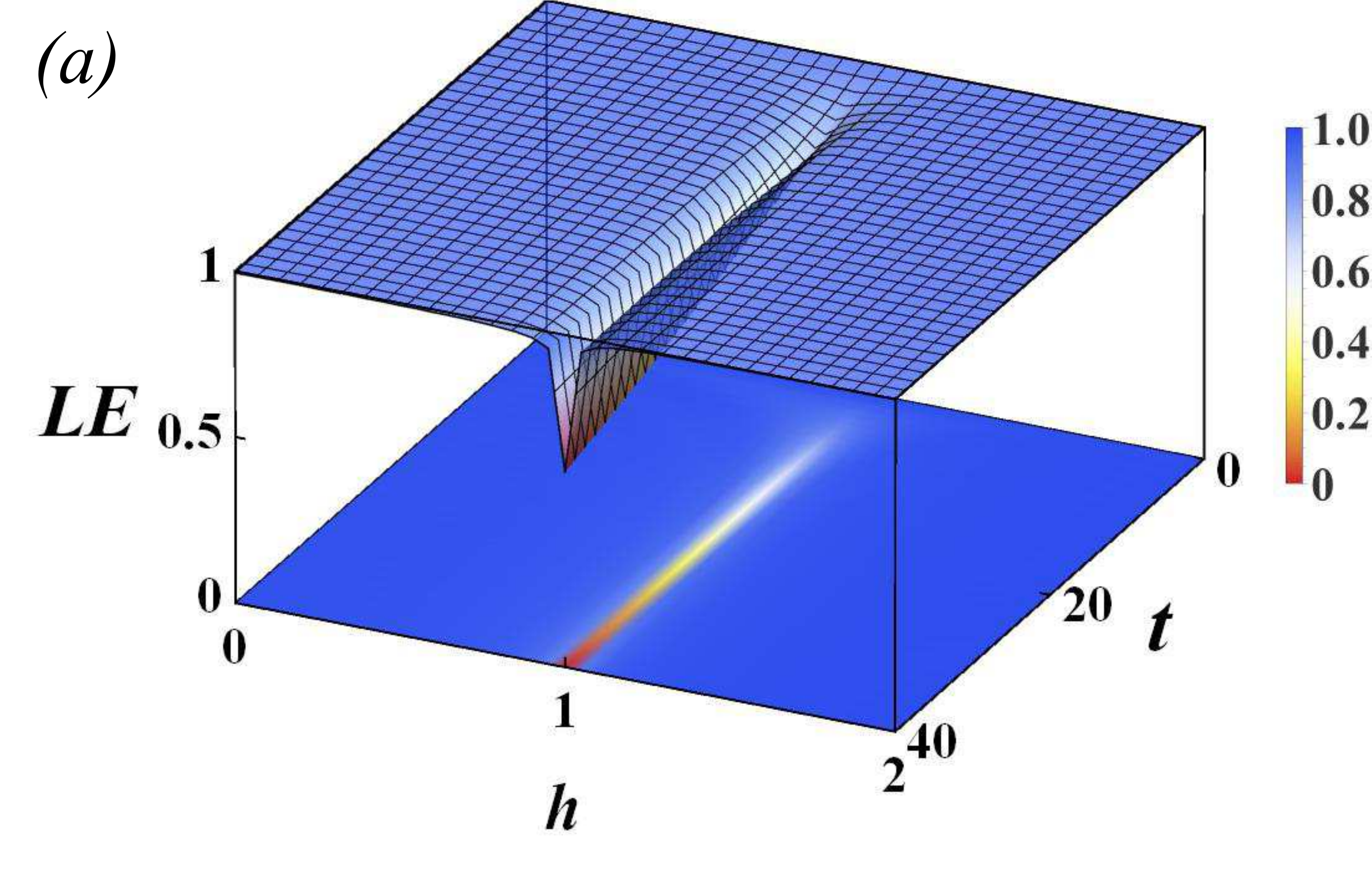}
\includegraphics[width=0.34\linewidth]{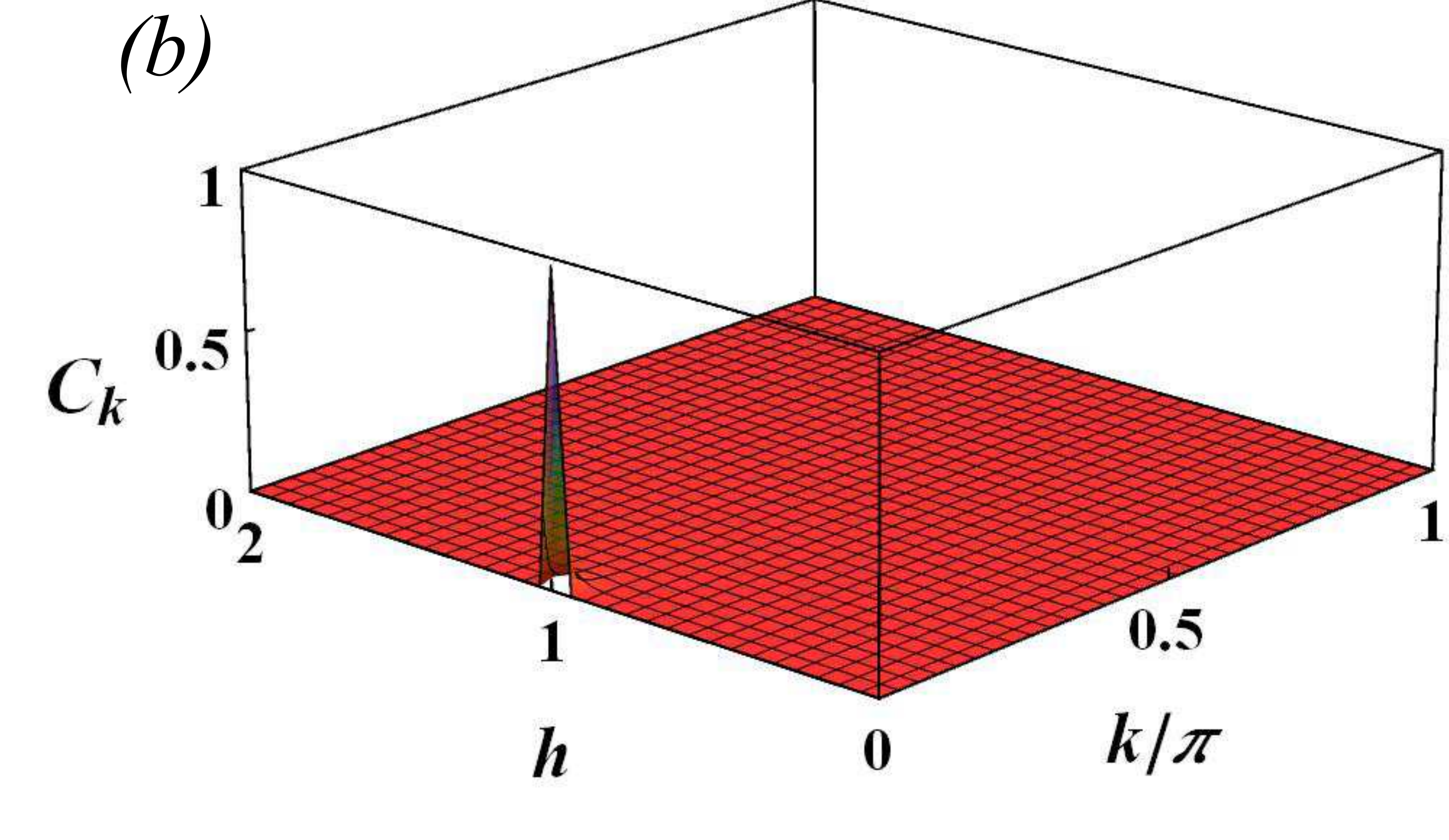}
\includegraphics[width=0.28\linewidth]{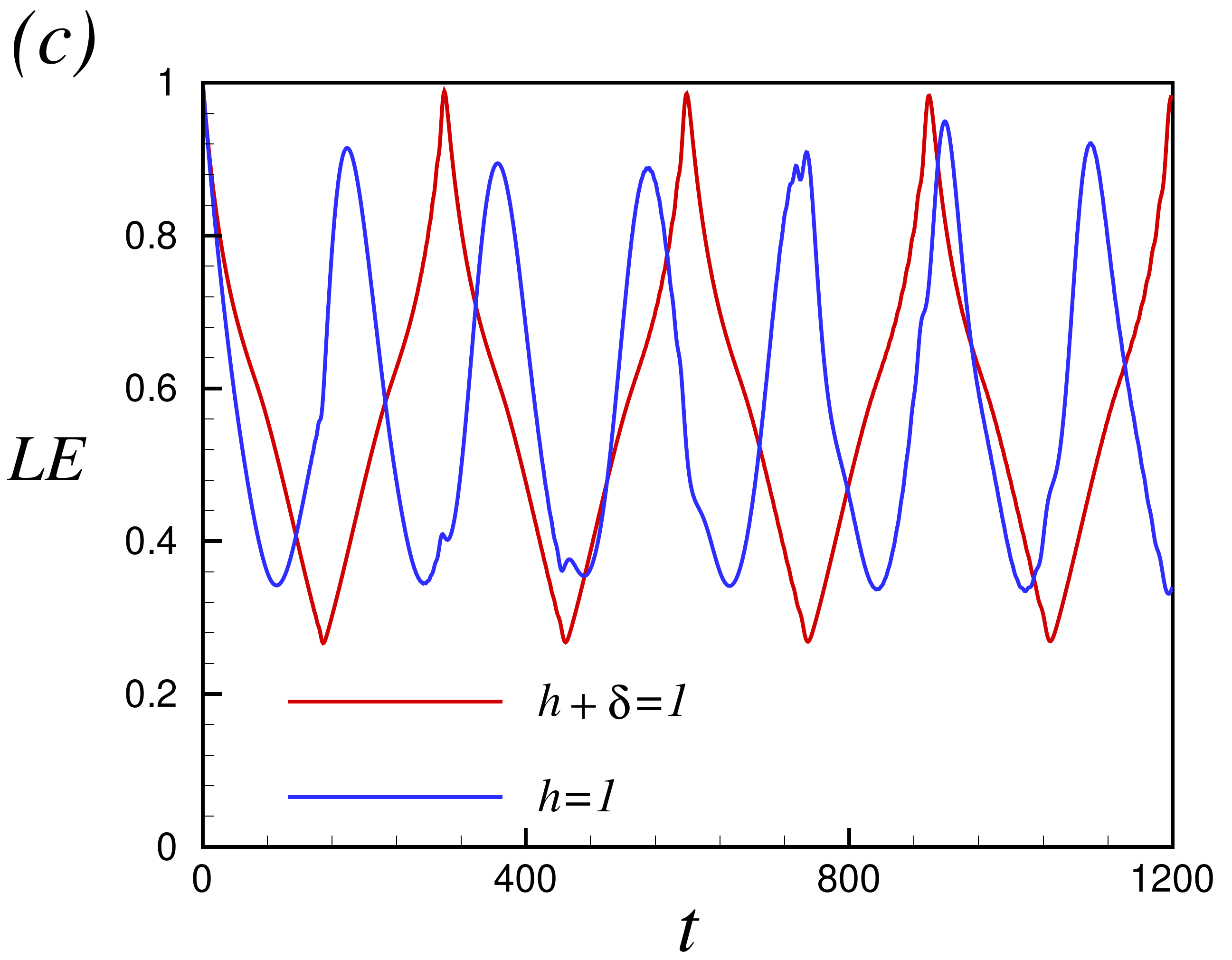}}
\caption{ (Color online) (a) Three-dimensional plot of
the LE in Eq. (\ref{Loschmidt}) as function of time $t$ and magnetic field $h$.
(b) The oscillation amplitude $C_{k}$ in the mode decomposition of the LE, Eq. (\ref{eq9}), as function of crystal momentum $k$
and magnetic field $h$.
(c) The LE as function of time $t$ at the critical point of the
unperturbed (perturbed) QCC with magnetic field $h=1 \ (h=1-\delta$).
The Hamiltonian parameters in all three panels are set to $J_{o}\!=\!1,
J_e\!=\!2$, $\theta/\pi=1/4$, $\delta=0.01$ and $N=400$.}
\label{fig6}
\end{figure*}
%
The essential role of the quasiparticles and their excitations in driving the behavior of the Loschmidt echo $-$ and the associated decoherence of the coupled qubit $-$ should now be clear. As detailed above, the quasiparticles play a double role. First, their excitations may restructure the unperturbed eigenstates of $H_k$ in (\ref{commH}) substantially when prevalent, making possible large state overlaps and, by that, large oscillation amplitudes in the mode decomposition of the LE. Secondly, the curvature of the quasiparticle bands determine the revival structure of the LE. A large/small curvature with a resulting large/small group velocity $v_g$ of the quasiparticles will set the time scale on which the qubit dynamics appears to be Markovian.

\subsection{Loschmidt echo: finite magnetic field}

The unperturbed QCC in a magnetic field $h$ exhibits a critical line $h_c\!=\! \pm\cos(\theta)\sqrt{J_{o}J_{e}}$ parameterized by $\theta, J_o$ and $J_e$ \cite{You2014a}.
Choosing $J_{o}=1$, $J_{e}=2$, $\theta=\pi/4$, $\delta=0.01$ and $N=400$, we have plotted the corresponding LE versus $h$ and $t$ in FIG. (\ref{fig6})(a).
As expected, the LE shows a single dip at the critical field $h_{c}=1$. This result is generic: With $\theta, J_e$, and $J_o$ fixed,
the LE suffers an enhanced decay only at the corresponding critical field of the QCC Hamiltonian, be it unperturbed ($h_c\!=\! \pm\cos(\theta)\sqrt{J_{o}J_{e}}$) or perturbed
($h_c\!=\! \delta\pm\cos(\theta)\sqrt{J_{o}J_{e}}$). In both cases the revival time of the LE is controlled by the group velocity of the
quasiparticles in the $\varepsilon_{k}^{(2)}$ band of the perturbed Hamiltonian. The magnetic field bends this band (cf. FIG. (\ref{fig5})(c)), and as a result the group velocities can be significantly larger
than in the case  when the field is zero. Moreover, numerical computations show that all oscillation amplitudes are very small in parameter space
except $C_{k}$ which takes a large value at the critical field in the center of the Brillouin zone (FIG. (\ref{fig6})(b)).  Putting these facts together, we
expect that the time evolution of the LE manifests distinct decays and revivals at the critical field.
This is verified In FIG. (\ref{fig6})(c) where the LE has been plotted versus time for $h_{c}=1$ (unperturbed QCC, blue curve)
and for $h_c\!=\!1-\delta$ (perturbed QCC, red curve). In both cases, the LE indeed exhibits deep valleys and high peaks,
however with different revival periods.
%
%
\section{Decoherence of a qubit in an extended-XY-model environment}
In this section we investigate the decoherence of a qubit embedded in an environment described by the one-dimensional extended $XY$ model with
a transverse staggered magnetic field \cite{Titvinidze2003}. While much of the methodology can be carried over from Secs. II and III, replacing the quantum compass chain
by the extended XY model will provide a complementary vista, adding to the picture of qubit decoherence in an interacting spin environment.

Imposing periodic boundary conditions, and assuming that the coupling to the qubit contains both a uniform ($\sim \delta$)
and a staggered ($\sim (-1)^n \delta_s$) component, the Hamiltonian of the composite system takes the form $H=H_{\text{env}} + H_{\text{q}} + H_{\text{int}}$, where
\bea
\label{eq11} \no
H_{\text{env}}&\!=\!&-\frac{1}{2} \sum_{n=1}^N \Big(\frac{J}{2}(\sigma_n^{x}\sigma_{n+1}^{x}+\sigma_{n}^{y}\sigma_{n+1}^{y})\\
&\!+\!&\frac{J_{3}}{4}(\sigma_{n}^{x}\sigma_{n+2}^{x}\!+\!\sigma_{n}^{y}\sigma_{n+2}^{y})\sigma_{n+1}^{z}\!+\!(-1)^{n}h_{s}\sigma_{n}^{z}\Big),\\
\no
H_{\text{q}}&\!=\!&\omega_{e}|e\rangle\langle e|, \ H_{\text{int}}\!=\!-\frac{1}{2}\Big(\delta\!+\!(-1)^{n}\delta_{s}|e\rangle\langle e|\Big)\sum_{n=1}^N\sigma_{n}^{z}.
\eea
We have here used the same tags for the Hamiltonians as in Sec. II, with $H_{\text{env}}$ and $H_{\text{q}}$ denoting the decoupled Hamiltonian of the environment and the qubit, respectively, and with
${H}_{\text{int}}$ the Hamiltonian of the qubit-environment interaction. Here $N$ counts the number of sites on the one-dimensional lattice,
$h_{s}$ is the magnitude of the staggered transverse magnetic field, and $J$ and $J_{3}$ are exchange couplings between spins on nearest-neighbor and next-nearest-neighbor sites, respectively. For simplicity we here consider the XX-limit of the model, with identical couplings in the $x$- and $y$-directions.

As in Sec. II we assume that the qubit is initially disentangled from the environment. In other words, the state $|\psi(0)\rangle$ of the composite system at time $t\!=\!0$ is given by
$|\psi(0)\rangle=|\phi_{\text{q}}(0)\rangle\otimes|\phi_{\text{env}}(0)\rangle$, with the normalized qubit state $|\phi_{\text{q}}(0)\rangle=c_{g}|g\rangle+c_{e}|e\rangle$ a superposition                                            of the ground state $|g\rangle$ and excited state $|e\rangle$, and where $|\phi_{\text{env}}(0)\rangle$ is the initial state of the environment.
With $U(t)=\exp (-i {H} t)$ the time-evolution operator, the time-evolved composite state can be written as
\bea
\label{eq12}
|\psi(t)\rangle&=&c_{g}|g\rangle \otimes \exp(-i {H}_{\text{env}}t)|\phi_{\text{env}}(0)\rangle\\
\no
&+& \exp(-i \omega_{e}t)c_{e}|e\rangle \otimes \exp(-i {H}^{(\delta,\delta_{s})}_{\text{env}}t)|\phi_{\text{env}}(0)\rangle,
\eea
using that $[{H}_{\text{q}},{H}_{\text{env}}] = [{H}_{\text{q}},{H}_{\text{int}}]=0$. Here
\begin{equation}
\label{XYpert}
{H}^{(\delta,\delta_{s})}_{\text{env}}={H}_{\text{env}}+V_{\text{env}}(\delta, \delta_{s})
\end{equation}
is the perturbed Hamiltonian of the environment, with $V_{\text{env}}(\delta, \delta_{s})=-\frac{1}{2}[\delta+(-1)^{n}\delta_{s}]\sum_{n=1}^N\sigma_{n}^{z}$ the effective potential from the interaction with the qubit. Note that the perturbed Hamiltonian ${H}^{(\delta,\delta_{s})}_{\text{env}}$ describes the extended $XY$ model in a staggered transverse magnetic field $h_s + \delta_s$, with an added uniform transverse field $\delta$.

In order to investigate the decoherence process induced by the environment, we follow the same route as in Sec. II. Eq. (\ref{eq12}) implies that the reduced density matrix of the qubit takes the form
\bea
\label{eq13}
{\cal\rho}_{\text{q}}&=&Tr_{\text{env}}|\psi(t)\rangle \langle\psi(t)|=c_{g}^{2}|g\rangle\langle g|+c_{e}^{2}|e\rangle\langle e|\\
\no
&+&e^{-i \omega_{e}t}c^{\ast}_{g}c_{e}\nu(t)|e\rangle\langle g|+e^{i \omega_{e}t}c_{g}c^{\ast}_{e}\nu^{\ast}(t)|g\rangle\langle e|,
\eea
with $\nu(t)\!=\!\langle\phi_{\text{env}}(0)|\exp(i {H}_{\text{env}}^{(\delta, \delta_{s})}t)\exp(-i {H}_{\text{env}}t)|\phi_{\text{env}}(0)\rangle$ the decoherence factor, implying the LE
 ${\cal L}=|\nu(t)|^2$ \cite{Quan2006, Cucchietti2003}. Thus, as in our analysis of the QCC-induced decoherence of the qubit in Sec. III, the problem boils down to computing the
 LE of the environment, now described by the extended XY model in a staggered magnetic field, perturbed by the qubit.

\section{Loschmidt echo of the extended XY model \label{ESEXY}}

\subsection{Preliminaries}

To derive a closed form of the LE we must first diagonalize the unperturbed as well as the perturbed environmental Hamiltonian.
In fact, it is sufficient to diagonalize the perturbed Hamiltonian in (\ref{XYpert}), ${H}^{(\delta,\delta_{s})}_{\text{env}}$, since it reduces to the
unperturbed one, ${H}_{\text{env}}$ in (\ref{eq11}), by setting $\delta=\delta_s=0$.
As a first step we again exploit the Jordan-Wigner transformation (\ref{JW}),
and map ${H}^{(\delta,\delta_{s})}_{\text{env}}$ onto a free fermion model,
\bea
\nonumber
{H}_{\text{env}}^{(\delta, \delta_{s})}&=&-\frac{1}{2}\sum_{n=1}^{N}\Big(J(c_{n}^{\dagger}c_{n+1}+c^{\dagger}_{n+1}c_{n})\\
\nonumber
&+&\frac{J_{3}}{2}(c_{n}^{\dagger}c_{n+2}+c^{\dagger}_{n+2}c_{n})\\
\label{eq14}
&+&\Big[\delta+(-1)^{n}(h_{s}+\delta_{s})\Big](2c^{\dagger}_{n}c_{n}-1)\Big).
\eea

By introducing two independent fermions at each unit cell of the lattice, $c_{n}^{A}\equiv c_{2n-1}$ and $c_{n}^{B}\equiv c_{2n}$
and performing a Fourier transformation, one obtains
{\small
\begin{multline}
{H}_{\text{env}}^{(\delta, \delta_{s})}=\sum_{k}\Big(\epsilon^{A}(k)c_{k}^{A\dagger}c_{k}^{A}+\epsilon^{B}(k)c_{k}^{B\dagger}c_{k}^{B} \\
+\epsilon^{AB}(k)(c_{k}^{A\dagger}c_{k}^{B} +c_{k}^{B\dagger}c_{k}^{A})\Big),
\label{eq15}
\end{multline}
}
where
\bea
\label{epsilonA}
\epsilon^{A}(k)&\!=\!&\frac{J_{3}}{2}\cos(k)\!-\!\delta\!+\!(h_s\!+\!\delta_{s}),\\
\label{epsilonB}
\epsilon^{B}(k)&\!=\!&\frac{J_{3}}{2}\cos(k)\!-\!\delta-(h_{s}\!+\!\delta_{s}), \\
\label{epsilonAB}
\epsilon^{AB}(k)&\!=\!&-J\cos(k/2),
\eea
and $k=4\pi n/N$ with $-N/4\le n \le N/4$ \cite{Divakaran2013}.
Using the Bogoliubov-type transformation
\bea
\no
c_{k}^{A}&=& \cos(\theta_{k}^{(\delta_{s})}/2) \alpha_{k}+\sin(\theta_{k}^{(\delta_{s})}/2) \beta_{k},\\
\no
c_{k}^{B}&=& -\sin(\theta_{k}^{(\delta_{s})}/2) \alpha_{k}+ \cos(\theta_{k}^{(\delta_{s})}/2) \beta_{k},
\eea
where
\bea
\no
\theta_{k}^{(\delta_{s})}
=-\arctan(J\cos(k/2)/(h_{s}+\delta_{s})),
\label{eq16}
\eea
we can finally write the Hamiltonian on diagonal form,
${H}_{\text{env}}^{(\delta, \delta_{s})}\!=\!\sum_{k}[\varepsilon^{\alpha}_{k}(\delta,\delta_{s})\alpha^{\dagger}_{k} \alpha_{k}+\varepsilon^{\beta}_{k}(\delta,\delta_{s})\beta^{\dagger}_{k}\beta_{k}]$,
%
%
with
\bea
\no
\varepsilon^{\alpha}_{k}(\delta,\delta_{s})\!=\!(J_{3}/2)\cos(k)\!-\!\delta\!-\!\sqrt{(h_{s}\!+\!\delta_{s})^{2}\!+\!J^{2}\cos^{2}(k/2)},\\
\no
\varepsilon^{\beta}_{k}(\delta,\delta_{s})\!=\!(J_{3}/2)\cos(k)\!-\!\delta\!+\!\sqrt{(h_{s}\!+\!\delta_{s})^{2}\!+\!J^{2}\cos^{2}(k/2)}.
\eea
The corresponding quasiparticle eigenstates are given by
%
\bea
\label{eq17}
\no
\alpha^{(\delta,\delta_{s})\dagger}_{k}|V\rangle&=&\cos(\theta_{k}^{(\delta_{s})}/2)c_{k}^{A\dagger}|0\rangle
-\sin(\theta_{k}^{(\delta_{s})}/2)c_{k}^{B\dagger}|0\rangle,\\
\no
\beta^{(\delta,\delta_{s})\dagger}_{k}|V\rangle&=&\sin(\theta_{k}^{(\delta_{s})}/2)c_{k}^{A\dagger}|0\rangle
+\cos(\theta_{k}^{(\delta_{s})}/2)c_{k}^{B\dagger}|0\rangle,
\eea
%
where $|V\rangle$ and $|0\rangle$ are vacuum states of the quasiparticle and fermion, respectively.
Notably, the quasiparticle operators of the unperturbed Hamiltonian, ($\alpha^{(0)}_{k}, \beta^{(0)}_{k}$), can be expressed on closed form as a linear combination
of those of the perturbed Hamiltonian, ($\alpha^{(\delta,\delta_{s})}_{k},\beta^{(\delta,\delta_{s})}_{k}$),
\bea
\label{eq18}
\no
\alpha^{(0)}_{k}&=&\cos(\eta_{k})\alpha^{(\delta,\delta_{s})}_{k}
-\sin(\eta_{k})\beta^{(\delta,\delta_{s})}_{k},\\
\no
\beta^{(0)}_{k}&=&\sin(\eta_{k})\alpha^{(\delta,\delta_{s})}_{k}
+\cos(\eta_{k})\beta^{(\delta,\delta_{s})}_{k}
\eea
where $2\eta_{k}=\theta_{k}^{(0)}-\theta_{k}^{(\delta_{s})}$.
It follows that eigenstates of the unperturbed Hamiltonian can be written in terms of the eigenstates of the perturbed Hamiltonian as
\bea
\label{eq19a}
\alpha^{(0)\dagger}_{k}|V\rangle&\!=\!&\cos(\eta_{k})\alpha^{(\delta,\delta_{s})\dagger}_{k}|V\rangle
\!-\!\sin(\eta_{k})\beta^{(\delta,\delta_{s})\dagger}_{k}|V\rangle,\\
\label{eq19b}
\beta^{(0)\dagger}_{k}|V\rangle&\!=\!&\sin(\eta_{k})\alpha^{(\delta,\delta_{s})\dagger}_{k}|V\rangle
\!+\!\cos(\eta_{k})\beta^{(\delta,\delta_{s})\dagger}_{k}|V\rangle.
\eea
The relations in Eqs. (\ref{eq19a}) and (\ref{eq19b}) will turn out to be useful when calculating the LE (next subsection).
But before turning to that task, let us briefly summarize what is known about the phase diagram of the (unperturbed) extended XY model in a transverse magnetic field.

The problem has been investigated comprehensively in Ref. [\onlinecite{Titvinidze2003}], revealing three
phases: one long-range-ordered antiferromagnetic phase and two distinct spin-liquid phases, denoted spin liquid (I) and spin liquid (II), respectively. The QPT between the antiferromagnetic phase and spin liquid (I) is a gapped-to-gapless transition which occurs at critical staggered fields $h_{s}^{c1}\!=\!\pm J_{3}/2$. The system is in the antiferromagnetic phase for $|h_{s}|\!\geq \!J_{3}/2$ where $\varepsilon^{\alpha}_{k}(0)\leqslant0$ and $\varepsilon^{\beta}_{k}(0)\!>\!0$
for all $k$ modes, and accordingly the ground state $|G_{\text{AFM}}\rangle$ takes the form $|G_{\text{AFM}}\rangle \sim \prod_k \alpha_k^{(0)\dagger}|V\rangle$ with energy $E_{\text{AFM}}=\sum_{k}\varepsilon^{\alpha}_{k}(0)$. When $\sqrt{J_{3}^{2}/4-1}<|h_{s}|<J_{3}/2$, the system enters spin liquid phase (I) where again
$\varepsilon^{\alpha}_{k}(0)\leqslant0$ for all $k$ modes, but now with $\varepsilon^{\beta}_{k}(0)$ also being negative for {\em some} $k$ modes.
Thus, the spin liquid (I) ground state takes the form $|G_{\text{(I)}}\rangle \sim \prod_{k,k^{\prime}} \alpha_k^{(0)\dagger} \beta_{k^{\prime}}^{(0)\dagger} |V\rangle$, with $k^{\prime}$ indexing those $\beta$-modes which have negative energies. At the critical points $h_{s}^{c2}=\pm\sqrt{J_{3}^{2}/4-1}$, a gapless-gapless QPT takes place between the spin liquid (I) and (II) phases, with a concurrent change of the Fermi surface topology \cite{Titvinidze2003}. In spin liquid phase (II), with $|h_{s}|\leq\sqrt{J_{3}^{2}/4-1}$, both $\varepsilon^{\alpha}_{k}(0)$ and $\varepsilon^{\beta}_{k}(0)$ have positive {\em and} negative branches, resulting in four Fermi points, two from each branch. Consequently, the spin liquid (II) ground state can be written as $|G_{\text{(II)}}\rangle \sim \prod_{k,k^{\prime}} \alpha_k^{(0)\dagger} \beta_{k^{\prime}}^{(0)\dagger} |V\rangle$, with $k$ and $k^{\prime}$ indexing the negative-energy $\alpha$- and $\beta$-modes, respectively.
%
\begin{figure*}
\centerline{\includegraphics[width=0.36\linewidth]{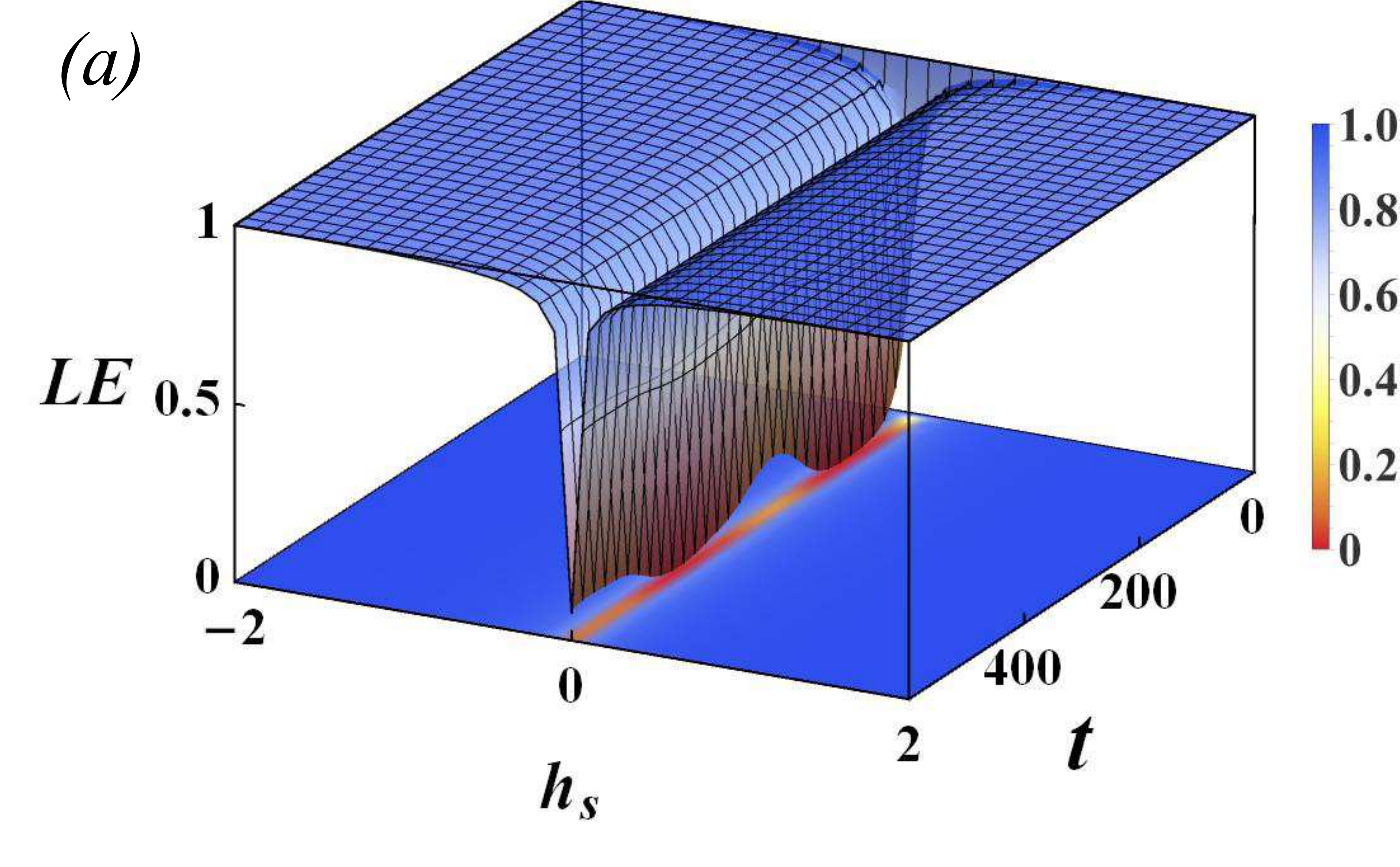}
\includegraphics[width=0.33\linewidth]{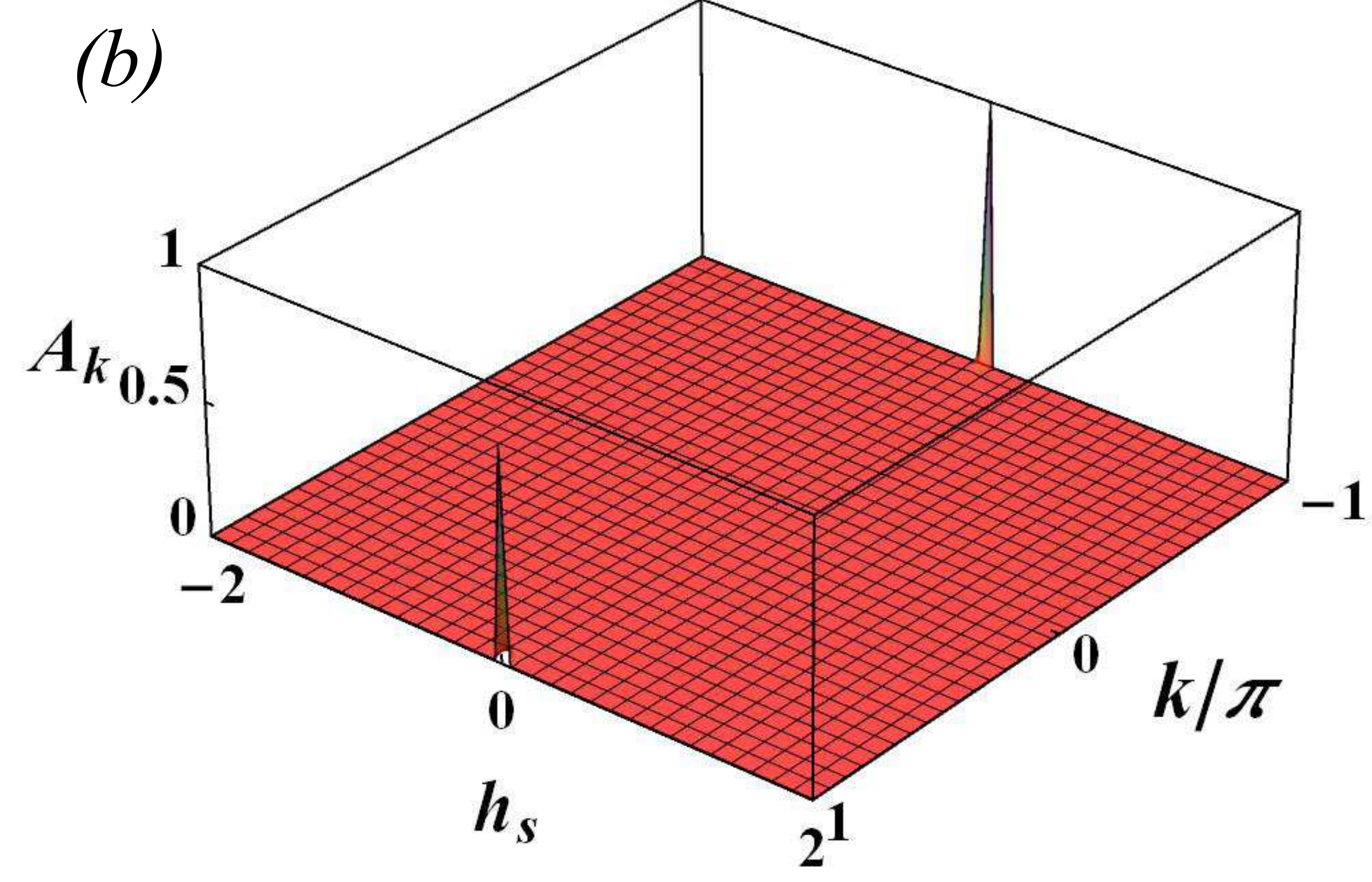}
\includegraphics[width=0.28\linewidth]{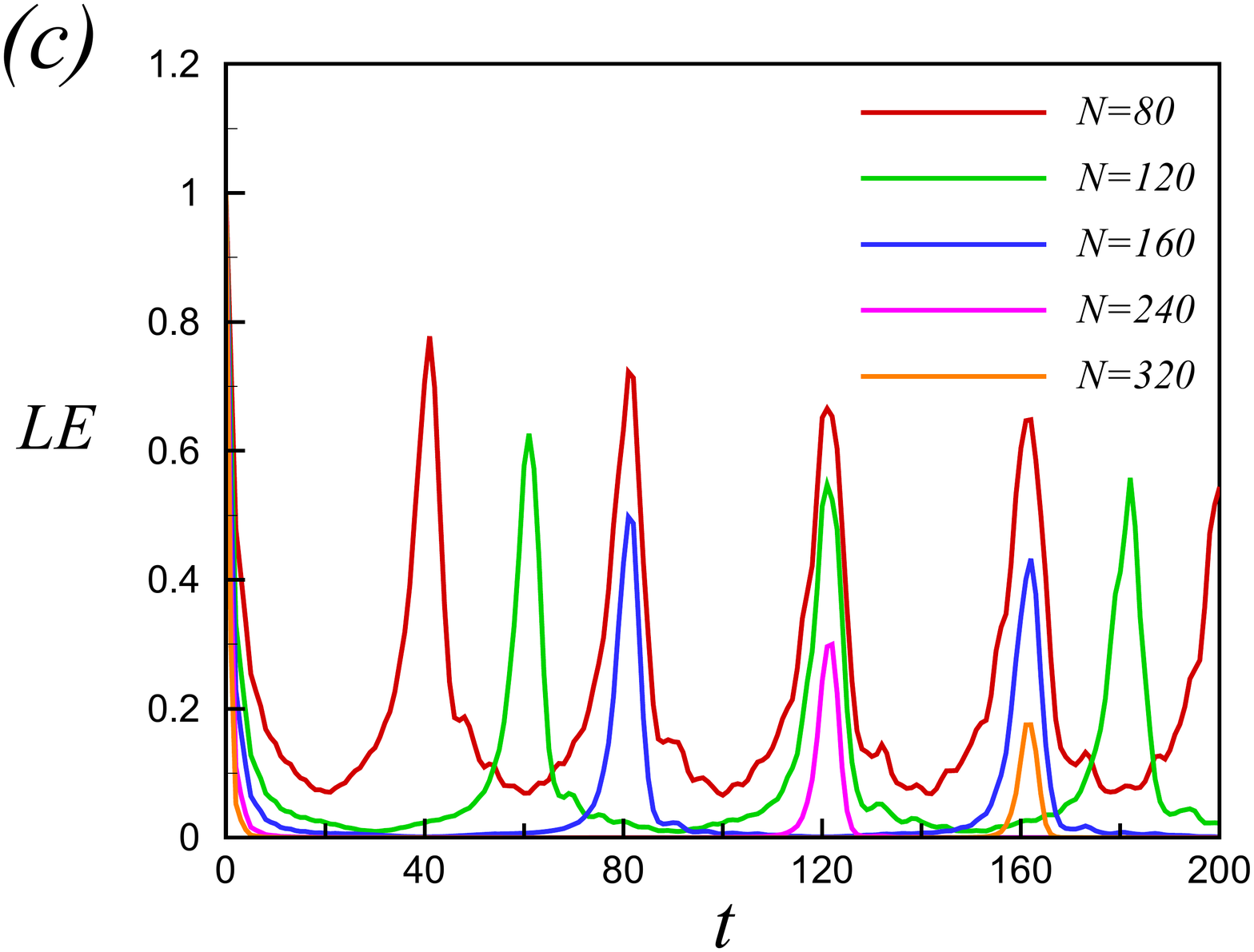}}
\caption{ (Color online) (a) Three-dimensional plot of
the LE in Eq. (\ref{XYLoschmidt}) as function of time $t$ and staggered magnetic field $h_s$
for $J=1$, $J_{3}=4$, $\delta_{s}=0.01$, and $N=1200$. (b) The oscillation amplitude $A_{k}$
in the mode decomposition of the LE, Eq. (\ref{eq20}), as function of crystal momentum $k$
and staggered magnetic field $h_s$, with the same parameter values as in (a).
(c) The LE, Eq. (\ref{XYLoschmidt}), for different system sizes $N$ versus time $t$
for $J_3=4$, $h_{s}=0.0$ and $\delta_s=0.1$.}
\label{fig7}
\end{figure*}
%

\subsection{Loschmidt echo: quantum-classical transitions at noncritical points \label{QCT}}

We now turn to the calculation of the LE. To be specific we may assume that the environment is initially prepared in the
antiferromagnetic ground state, with parameters chosen to put it close to the phase transition to the spin liquid phase (I),
\begin{equation}
\label{AFM}
|\phi_{\text{env}}(0)\rangle=\prod_{-\pi \le k \le \pi}\alpha^{(0)\dagger}_{k}|V\rangle.
\end{equation}
This choice of initial environmental state allows us to probe the LE at criticality by using ${H}_{\text{env}}^{(\delta, \delta_{s})}$ to do a quantum quench
to one of the critical points $h_{s}^{c1}\!=\!\pm J_{3}/2$. To explore the full spin liquid (I) phase away from criticality one instead choses the ground state
\begin{equation}
\label{spinliquidI}
|\phi_{\text{env}}(0)\rangle=\prod_{k,k^{\prime}}\alpha^{(0)\dagger}_{k} \beta^{(0)\dagger}_{k^{\prime}}|V\rangle,
\end{equation}
with $-\pi \le k \le \pi$ and $ \varepsilon^{\beta}_{k}(0) \le 0$, as initial environmental state.
Injecting Eqs. (\ref{eq19a}) and (\ref{eq19b}) into (\ref{AFM}) or (\ref{spinliquidI}) and using the expression for the LE,
\begin{equation} \label{XYLoschmidt}
{\cal L}(t) = |\langle\phi_{\text{env}}(0)|\exp(i {H}_{\text{env}}^{(\delta, \delta_{s})}t)\exp(-i {H}_{\text{env}}t)|\phi_{\text{env}}(0)\rangle|^2,
\end{equation}
it is straightforward to show that in both cases the LE reduces to the form
\bea
\label{eq20}
{\cal L}(t)=\prod_{-\pi \le k \le \pi}|1-A_{k}\sin^{2}(\frac{\Delta\varepsilon_{k}t}{2})|
\eea
where
\bea
\no
A_{k}&&=\sin^{2}(2\eta_{k}), \\
\Delta\varepsilon_{k}&&=2\sqrt{(h_{s}+\delta_{s})^{2}+J^{2}\cos^{2}(k/2)}.
\label{eq21}
\eea
By inspection, neither $A_k$ nor $\Delta\varepsilon_{k}$ depend on
$\delta$. It follows from (\ref{eq20}) that in the case when the qubit-environment interaction only contains a uniform coupling $\sim \delta$,
with $\delta_s =0$, ${\cal L}(t) = 1$ independent of the strength of $\delta$. As an upshot, the state of a qubit embedded in a spin environment
here described by an extended XY model in a transverse staggered magnetic field does {\em not} decohere as long as the staggered interaction component vanishes,
regardless of the strength of the uniform qubit-environment interaction. This result, similarly uncovered for a central spin model with the qubit coupled to an ordinary XY
chain \cite{Yuan2007a}, may suggest practical strategies for protecting qubits in
applications for quantum information technologies.

In FIG. \ref{fig7}(a) we have plotted the LE of the environment perturbed by the qubit with a staggered interaction $\sim \delta_s = 0.01$ as a function of staggered magnetic field $h_{s}$ and time $t$.
As seen in the figure, the LE displays neither enhanced decays nor revival structures at the critical points $h_{s}^{c1}=\pm2$ or $h_{s}^{c2}=\pm\sqrt{3}$ of the environment as opposed to what reports in previous works \cite{Quan2006,Yuan2007a}. Instead it shows an
accelerated decay at $h_{s}=0.$ The point $h_{s}=0$, while being a critical point of
the extended XY model in the absence of three-site spin interaction (i.e. with $J_{3}=0$ in (\ref{eq11})), is noncritical for any nonzero value of $J_{3}$.
{\em But how can the LE exhibit an accelerated decay at a non-critical value of the staggered field? And why does the LE not exhibit an accelerated decay when the staggered field is critical?}

To answer these questions we emulate our analysis from Sec. III.  A numerical check confirms that the absence of an accelerated decay of the LE along the critical lines $h_s \!=\! \pm J_3/2$ comes about because of the small values of the oscillation amplitudes $A_k$ in (\ref{eq20}) for all $k$. In exact analogy to the critical QCC away from the IP, the smallness of the $A_k$-amplitudes is a consequence of the fact that the quasiparticles which control the LE remain gapped at criticality. On the contrary, the accelerated decays of the LE which are manifest in both environmental models $-$ the QCC and the extended XY model $-$ are correlated with large oscillation amplitudes in the LE mode decompositions, Eq. (\ref{eq8}) and (\ref{eq20}), respectively. As we have seen, large oscillation amplitudes are favored by the presence of easily excited quasiparticles. Importantly, not only may a quantum phase transition not favor LE-controlling quasiparticle excitations, but such excitations may instead appear {\em within} a stable phase, such as the type-I spin liquid phase of the extended XY model. This can be confirmed numerically. In FIG. (\ref{fig7})(b) we display the oscillation amplitude $A_k$ versus $k$ and $h_{s}$, with Hamiltonian parameters $J_{3}=4$, $\delta_{s}=0.01$, and $N=1200$. It is clearly seen that $A_k$ vanishes everywhere except in the neighborhood of $h_{s}=0$ at the Brillouin boundary zone boundary where the extended XY model becomes massless, with propagating quasiparticles \cite{Titvinidze2003}. In FIG. \ref{fig7}(c) we have computed the time-dependence of the LE for different system sizes,
 verifying that the LE revivals get attenuated, with longer periods, as the system gets larger.

\section{Summary}
Based on two case studies of a qubit coupled to an interacting spin environment -- with the environment modeled by a quantum compass chain or an extended XY model in a transverse staggered magnetic field -- we arrive at the conclusion that the presence of a quantum phase transition is neither a sufficient nor a necessary condition for an accelerated decoherence rate of the qubit.
By examining how the eigenstates of the models imprint the Loschmidt echo -- and by that the decay rate of the qubit -- we find that what {\em does} matter is the availability of propagating quasiparticles which couple to the qubit via a back action (as signaled by their having an impact on the Loschmidt echo). While a quantum phase transition generically supports massless excitations, our case study of the QCC reveals that these excitations may not necessarily couple to the qubit, and therefore do not influence its decoherence rate.  {\em This observation invalidates the conventional view that the closeness of an environment to a quantum phase transition is inherently linked to an enhanced decoherence of a system embedded in it.} Taking the extended XY model as environmental model, the quasiparticles in one of its spin-liquid phases are found to couple to the qubit. This provides an example that a stable massless phase can act as a source of accelerated decoherence. Our findings may prove useful when developing strategies to reduce decoherence in quantum devices with interacting qubits.

\begin{acknowledgments}
This research was supported by the Swedish Research Council (Grant No. 621-2014-5972). \\ \\
\end{acknowledgments}

\section{Appendix}

The amplitudes in the mode decomposition of the Loschmidt echo, Eq. (\ref{eq9}), depend on the state overlaps $F_{m,k}=|\langle\psi_{m,k}(\delta)|\psi_{0,k}(0)\rangle|^{2}$ $(m=0,...,7)$ as
\bea
\label{eqAC2}
\no
A_{0,k}&=&4F_{0,k}F_{7,k},\\
\no
B_{0,k}&=&4(F_{2,k}+F_{3,k}+F_{4,k}+F_{5,k})(F_{0,k}+F_{7,k}),\\
\no
A_{1,k}&=&4F_{1,k}F_{6,k},\\
\no
B_{1,k}&=&4(F_{2,k}+F_{3,k}+F_{4,k}+F_{5,k})(F_{1,k}+F_{6,k}),\\
\no
C_{k}&=&4(F_{0,k}F_{1,k}+F_{6,k}F_{7,k}),\\
\no
D_{k}&=&4(F_{0,k}F_{6,k}+F_{1,k}F_{7,k}).
\eea
Here $|\psi_{m,k}(\delta)\rangle$ are eigenstates of the Hamiltonian $H_k$ in Eq. (\ref{commH}) with $h=\delta$.

At the critical line $\theta_c= \pi/2$ in $(\theta,J_e/J_o)$-space, the LE reduces to the simple form
\bea
\no
{\cal L}(\theta_{1},\theta_{c},t)=\!\prod_{0 \le k \le \pi}|1\!-\!A_{k}\sin^{2}(\varepsilon_{k}^{1}(\delta)t)\!-\!B_{k}\sin^{2}(\frac{\varepsilon_{k}^{1}(\delta)t}{2})|,
\eea \\
where $A_{k}=4(F_{0,k}+F_{1,k})(F_{6,k}+F_{7,k})$ and $B_{k}=4(F_{0,k}+F_{1,k}+F_{6,k}+F_{7,k})(F_{2,k}+F_{3,k}+F_{4,k}+F_{5,k})$, and where
$\varepsilon_{k}^{1}(\delta)$ are energies of the quasiparticles that fill up the lowest band of $H_{\text{env}}^{(\delta)}$ in Eq. (\ref{pertqcc}).

%
\bibliography{Ref}

%
%

\end{document}